\documentclass[rotating,dvips]{arxstspdf}
\usepackage{dcolumn}
\usepackage{graphicx}
\usepackage{stfloats,flushend}

\volume{25}
\issue{1}
\pubyear{2010}
\firstpage{51}
\lastpage{71}
\doi{10.1214/10-STS321}

\makeatletter
\newcolumntype{d}[1]{D{.}{.}{#1}}
\newcommand{\eqref}[1]{(\ref{#1})}
\DeclareMathAlphabet\mathcaligr{OMS}{cmsy}{m}{n}
\renewcommand{\mathcal}{\mathcaligr}
\newcommand\indep{\protect\mathpalette{\protect\independenT}{\perp}}

\newcommand{\E}{\mathbb{E}}
\newcommand{\bone}{\mathbf{1}}
\newcommand{\sgn}{\operatorname{sgn}}
\newcommand{\var}{\operatorname{Var}}
\newcommand{\cov}{\operatorname{Cov}}
\def\independenT#1#2{\mathrel{\rlap{$#1#2$}\mkern2mu{#1#2}}}

\renewcommand{\citet}[1]{(\citeyear{#1})}
\renewcommand{\citep}[1]{\citeyear{#1}}
\renewcommand{\epsilon}{\varepsilon}
\newtheorem{theorem}{Theorem}

\newproclaim{assumption}{Assumption}
\newtheorem{corollary}{Corollary}

\newproclaim{remark}{Remark}
\makeatother

\begin{document}
\begin{frontmatter}

\title{Identification, Inference and Sensitivity Analysis
for Causal Mediation Effects}
\runtitle{Causal Mediation Analysis}

\begin{aug}
\author[a]{\fnms{Kosuke} \snm{Imai}\corref{}\ead[label=e1]{kimai@princeton.edu}\ead[label=u1,url]{http://imai.princeton.edu}},
\author[b]{\fnms{Luke} \snm{Keele}\ead[label=e2]{keele.4@polisci.osu.edu}}
\and
\author[c]{\fnms{Teppei} \snm{Yamamoto}\ead[label=e3]{tyamamot@princeton.edu}}
\runauthor{K. Imai, L. Keele and T. Yamamoto}

\affiliation{Princeton University, Ohio State University and Princeton
University}

\address[a]{Kosuke Imai is Assistant Professor, Department of
Politics, Princeton University, Princeton, New Jersey 08544, USA
\printead{e1,u1}.}
\address[b]{Luke Keele is Assistant Professor, Department
Political Science, Ohio State University, 2140 Derby Hall,
Columbus, Ohio 43210, USA \printead{e2}.}
\address[c]{Teppei Yamamoto is Ph.D. Student, Department of Politics,
Princeton University,
031 Corwin Hall, Princeton, New Jersey 08544, USA \printead{e3}.}

\end{aug}

%
\begin{abstract}
Causal mediation analysis is routinely conducted by applied
researchers in a variety of disciplines. The goal of such an
analysis is to investigate alternative causal mechanisms by
examining the roles of intermediate variables that lie in the causal
paths between the treatment and outcome variables. In this paper
we first prove that under a particular version of sequential
ignorability assumption, the average causal mediation effect (ACME)
is nonparametrically identified. We compare our identification
assumption with those proposed in the literature. Some practical
implications of our identification result are also discussed. In
particular, the popular estimator based on the linear structural
equation model (LSEM) can be interpreted as an ACME estimator once
additional parametric assumptions are made. We show that these
assumptions can easily be relaxed within and outside of the LSEM
framework and propose simple nonparametric estimation strategies.
Second, and perhaps most importantly, we propose a new sensitivity
analysis that can be easily implemented by applied researchers
within the LSEM framework. Like the existing identifying
assumptions, the proposed sequential ignorability assumption may be
too strong in many applied settings. Thus, sensitivity analysis is
essential in order to examine the robustness of empirical findings
to the possible existence of an unmeasured confounder. Finally, we
apply the proposed methods to a randomized experiment from political
psychology. We also make easy-to-use software available to
implement the proposed methods.
\end{abstract}

%
\vspace*{-0.5pt}

\end{frontmatter}

\section{Introduction}

Causal mediation analysis is routinely conducted by applied
researchers in a variety of scientific disciplines including
epidemiology, political science, psychology and sociology
(see Mac{K}innon, \citep{mack08}). The goal  of such an analysis is to
investigate causal mechanisms by examining the role of intermediate
variables thought to lie in the causal path between the treatment and
outcome variables. Over fifty years ago, Cochran \citet{coch57} pointed to
both the possibility and difficulty of using covariance analysis to
explore causal mechanisms by stating: ``Sometimes these averages have
no physical or biological meaning of interest to the investigator, and
sometimes they do not have the meaning that is ascribed to them at
first glance'' (page 267). Recently, a number of statisticians have
taken up Cochran's challenge. Robins and Greenland \citet{robigree92} initiated a formal
study of causal mediation analysis, and a number of articles have
appeared in more recent years
(e.g., Pearl, \citep{pear01};
Robins, \citeyear{robi03};
Rubin, \citeyear{rubi04};
Petersen, Sinisi and van der Laan, \citeyear{petesinilaan06};
Geneletti, \citeyear{gene07};
Joffe,  Small and Hsu, \citeyear{joffetal07};
Ten Have et al., \citeyear{tenhetal07};
Albert, \citeyear{albe08};
Jo, \citeyear{jo08};
Joffe et al., \citeyear{joffetal08};
Sobel, \citeyear{sobe08};
VanderWeele, \citeyear{vand08a,vand09};
Glynn, \citeyear{glyn10}).

What do we mean by a causal mechanism? The aforementioned paper by
Cochran gives the following example. In a randomized experiment,
researchers study the causal effects of various soil fumigants on
eelworms that attack farm crops. They observe that these soil
fumigants increase oats yields but wish to know whether the reduction
of eelworms represents an intermediate phenomenon that mediates this
effect. In fact, many scientists across various disciplines are not
only interested in causal effects but also in causal mechanisms
because competing scientific theories often imply that different causal
paths underlie the same cause-effect relationship.

In this paper we contribute to this fast-growing literature in
several ways. After briefly describing our motivating example in the
next section, we prove in Section~\ref{sec:identification} that under
a particular version of the sequential ignorability assumption, the
average causal mediation effect (ACME) is nonparametrically
identified. We compare our identifying assumption with those proposed
in the literature, and discuss practical implications of our
identification result. In particular, Baron and Kenny's \citet{barokenn86} popular
estimator (Google Scholar records over 17 thousand citations for this
paper), which is based on a linear structural equation model (LSEM),
can be interpreted as an ACME estimator under the proposed assumption
if additional parametric assumptions are satisfied. We show that
these additional assumptions can be easily relaxed within and outside
of the LSEM framework. In particular, we propose a simple
nonparametric estimation strategy in Section~\ref{sec:estimation}. We
conduct a Monte Carlo experiment to investigate the finite-sample
performance of the proposed nonparametric estimator and its asymptotic
confidence interval.

\begin{table*}[b]
\tabcolsep=0pt
\caption{Descriptive statistics and estimated average treatment
effects from the
media framing experiment. The middle four columns show the means and standard
deviations of the mediator and outcome variables for each treatment
group. The
last column reports the estimated average causal effects of the public
order frame
as opposed to the free speech frame on the three response variables
along with
their standard errors. The estimates suggest that the treatment
affected each
of these variables in the expected directions} \label{tab:desc}
\begin{tabular*}{\textwidth}{@{\extracolsep{4in minus 4in}}lccccc@{}}
\hline
& \multicolumn{4}{c}{\textbf{Treatment media frames}}  \\
\cline{2-5}
& \multicolumn{2}{c}{\textbf{Public order}} &  \multicolumn{2}{c}{\textbf{Free
speech}}  \\
\ccline{2-3,4-5}
\textbf{Response variables} & \textbf{Mean} & \textbf{S.D.} &  \textbf{Mean} & \textbf{S.D.} & \textbf{ATE (s.e.)} \\
\hline
Importance of free speech & 5.25 & 1.43 &  5.49 & 1.35 & $-$0.231
(0.239) \\
Importance of public order & 5.43 & 1.73 &  4.75 & 1.80 &  \phantom{$-$}0.674
(0.303) \\
Tolerance for the KKK & 2.59 & 1.89 &  3.13 & 2.07 & $-$0.540 (0.340)
\\[3pt]
Sample size &\multicolumn{2}{c}{69} &  \multicolumn{2}{c}{67}  \\
\hline
\end{tabular*}
\end{table*}

Like many identification assumptions, the proposed assumption may be
too strong for the typical situations in which causal mediation
analysis is employed. For example, in experiments where the treatment
is randomized but the mediator is not, the ignorability of the
treatment assignment holds but the ignorability of the mediator may
not. In Section~\ref{sec:sens} we propose a new sensitivity analysis
that can be implemented by applied researchers within the standard
LSEM framework. This method directly evaluates the robustness of
empirical findings to the possible existence of unmeasured
pre-treatment variables that confound the relationship between the
mediator and the outcome. Given the fact that the sequential
ignorability assumption cannot be directly tested even in randomized
experiments, sensitivity analysis must play an essential role in
causal mediation analysis. Finally, in Section~\ref{sec:emp} we
apply the proposed methods to the empirical example, to which we now
turn.

\section{An Example from the Social Sciences}
\label{sec:example}

Since the influential article by Baron and Kenny \citet{barokenn86}, mediation
analysis has been frequently used in the social sciences and
psychology in particular. A~central goal of these disciplines is to
identify causal mechanisms underlying human behavior and opinion
formation. In a typical psychological experiment, researchers
randomly administer certain stimuli to subjects and compare treatment
group behavior or opinions with those in the control group. However,
to directly test psychological theories, estimating the causal effects
of the stimuli is typically not sufficient. Instead, researchers
choose to investigate psychological factors such as cognition and
emotion that mediate causal effects in order to explain why
individuals respond to a certain stimulus in a particular way.
Another difficulty faced by many researchers is their inability to
directly manipulate psychological constructs. It is in this context
that causal mediation analysis plays an essential role in social
science research.

In Section~\ref{sec:emp} we apply our methods to an influential
randomized experiment from political psychology.
Nelson, Clawson and
Oxley \citet{nelsclawoxle97} examine how the framing of political issues
by the news media affects citizens' political opinions. While the
authors are not the first to use causal mediation analysis in
political science, their study is one of the most well-known examples
in political psychology and also represents a typical application of
causal mediation analyses in the social sciences. Media framing is
the process by which news organizations define a political issue or
emphasize particular aspects of that issue. The authors hypothesize
that differing frames for the same news story alter citizens'
political tolerance by affecting more general political attitudes.
They conducted a randomized experiment to test this mediation
hypothesis.

Specifically, Nelson, Clawson and
Oxley \citet{nelsclawoxle97} used two different local newscasts
about a Ku Klux Klan rally held in central Ohio. In the
experiment, student subjects were randomly assigned to watch the
nightly news
from two different local news channels. The two news clips were
identical except for the final story on the Klan rally. In one
newscast, the Klan rally was presented as a free speech issue. In the
second newscast, the journalists presented the Klan rally as a
disruption of public order that threatened to turn violent. The outcome was
measured using two different scales of political tolerance. Immediately after
viewing the news broadcast, subjects were
asked two seven-point scale questions measuring their tolerance for
the Klan speeches and rallies. The hypothesis was that the causal
effects of the media frame on tolerance are mediated by subjects'
attitudes about the importance of free speech and the
maintenance of public order. In other words, the media frame
influences subjects' attitudes toward the Ku Klux Klan by encouraging
them to consider the Klan rally as an event relevant for the general
issue of free speech or public order. The researchers used additional
survey questions and a scaling method to measure these hypothesized
mediating factors after the experiment was conducted.

Table~\ref{tab:desc} reports descriptive statistics for these mediator
variables as well as the treatment and outcome variables. The sample
size is 136, with 67 subjects exposed to the free speech frame and 69
subjects assigned to the public order frame. As is clear from the
last column, the media frame treatment appears to influence both types
of response variables in the expected directions. For example, being
exposed to the public order frame as opposed to the free speech frame
significantly increased the subjects' perceived importance of public
order, while decreasing the importance of free speech (although the
latter effect is not statistically significant). Moreover, the public
order treatment decreased the subjects' tolerance toward the Ku Klux
Klan speech in the news clips compared to the free speech frame.

It is important to note that the researchers in this example are
primarily interested in the causal mechanism between media framing and
political tolerance rather than various causal effects given in the
last column of Table~\ref{tab:desc}. Indeed, in many social science
experiments, researchers' interest lies in the identification of
causal mediation effects rather than the total causal effect or
controlled direct effects (these terms are formally defined in the
next section). Causal mediation analysis is particularly appealing in
such situations.

One crucial limitation of this study, however, is that like many other
psychological experiments the original researchers were only able to
randomize news stories but not subjects' attitudes. This implies that
there is likely to be unobserved covariates that confound the
relationship between the mediator and the outcome. As we formally
show in the next section, the existence of such confounders represents
a violation of a key assumption for identifying the causal mechanism.
For example, it is possible that subjects' underlying political
ideology affects both their public order attitude and their tolerance
for the Klan rally within each treatment condition. This scenario is
of particular concern since it is well established that politically
conservative citizens tend to be more concerned about public order
issues and also, in some instances, be more sympathetic to groups like
the Klan. In Section~\ref{sec:sens} we propose a new sensitivity
analysis that partially addresses such concerns.

\section{Identification}
\label{sec:identification}

In this section we propose a new nonparametric identification
assumption for the ACME and discuss its practical implications. We
also compare the proposed assumption with those available in the
literature.

\subsection{The Framework}
\label{subsec:framework}

Consider a simple random sample of size $n$ from a population where
for each unit $i$ we observe $(T_i, M_i,\break  X_i, Y_i)$. We use $T_i$ to
denote the binary treatment variable where $T_i = 1$ ($T_i = 0$)
implies unit $i$ receives (does not receive) the treatment. The
mediating variable of interest, that is, the mediator, is represented by
$M_i$, whereas $Y_i$ represents the outcome variable. Finally, $X_i$
denotes the vector of observed pre-treatment covariates, and we use
$\mathcal{M}$, $\mathcal{X}$ and $\mathcal{Y}$ to denote the support
of the distributions of $M_i$, $X_i$ and $Y_i$, respectively.

What qualifies as a mediator? Since the mediator lies in the causal
path between the treatment and the outcome, it must be a
post-treatment variable that occurs before the outcome is realized.
Beyond this minimal requirement, what constitutes a mediator is
determined solely by the scientific theory under investigation.
Consider the following example, which is motivated by a referee's
comment. Suppose that the treatment is parents' decision to have
their child receive the live vaccine for H1N1 flu virus and the
outcome is whether the child develops flu or not. For a virologist, a
mediator of interest may be the development of antibodies to H1N1 live
vaccine. But, if parents sign a form acknowledging the risks of the
vaccine, can this act of form signing also be a mediator? Indeed,
social scientists (if not virologists!) may hypothesize that being
informed of the risks will make parents less likely to have their
child receive the second dose of the vaccine, thereby increasing the
risk of developing flu. This example highlights the important role of
scientific theories in causal mediation analysis.

To define the causal mediation effects, we use the potential outcomes
framework. Let $M_i(t)$ denote the potential value of the mediator
for unit $i$ under the treatment status $T_i = t$. Similarly, we use
$Y_i(t,m)$ to represent the potential outcome for unit $i$ when $T_i
= t$ and $M_i = m$.
Then, the observed variables can be written as $M_i = M_i(T_i)$ and
$Y_i = Y_i(T_i, M_i(T_i))$. Similarly, if the mediator takes $J$
different values, there exist $2J$ potential values of the outcome
variable, only one of which can be observed.

Using the potential outcomes notation, we can define the causal
mediation effect for unit $i$ under treatment status $t$ as
(see Robins and Greenland, \citep{robigree92}; Pearl, \citep{pear01})
%
\begin{equation}
\delta_i(t)  \equiv Y_i(t, M_i(1)) - Y_i(t, M_i(0)) \label{eq:def}
\end{equation}
for $t=0,1$. Pearl \citet{pear01} called $\delta_i(t)$ the \textit{natural
indirect effect}, while Robins \citet{robi03} used the term the \textit{pure
indirect effect} for $\delta_i(0)$ and the \textit{total indirect
effect} for $\delta_i(1)$. In words, $\delta_i(t)$ represents the
difference between the potential outcome that would result under
treatment status $t$, and the potential outcome that would occur if
the treatment status is the same and yet the mediator takes a value
that would result under the other treatment status. Note that the
former is observable (if the treatment variable is actually equal to
$t$), whereas the latter is by definition unobservable [under the
treatment status $t$ we never observe $M_i(1-t)$]. Some feel
uncomfortable with the idea of making inferences about quantities that
can never be observed (e.g., Rubin, \citep{rubi05}, page~325), while others
emphasize their importance in policy making and scientific
research (Pearl, \citeyear{pear01}, Section~2.4, \citeyear{pear10}, Section~6.1.4;
Hafeman and Schwartz \citeyear{hafeschw09}).

Furthermore, the above notation implicitly assumes that the potential
outcome depends only on the values of the treatment and mediating
variables and, in particular, not on \textit{how} they are realized. For
example, this assumption would be violated if the outcome variable
responded to the value of the mediator differently depending on
whether it was directly assigned or occurred as a natural response to
the treatment, that is, for $t=0,1$ and all $m \in\mathcal{M}$, $Y_i(t,
M_i(t))=Y_i(t, M_i(1-t))=Y_i(t,m)$ if $M_i(1)=M_i(0)=m$.

Thus, equation~\eqref{eq:def} formalizes the idea that the mediation
effects represent the indirect effects of the treatment through the
mediator. In this paper we focus on the identification and inference
of the average causal mediation effect (ACME), which is defined as
%
\begin{eqnarray}\label{eq:defbar}
 \bar\delta(t)  &\equiv& \E(\delta_i(t))\nonumber
\\[-8pt]
\\[-8pt]
  &=&  \E\{Y_i(t, M_i(1))- Y_i(t, M_i(0))\} \nonumber
\end{eqnarray}
for $t=0,1$. In the potential outcomes framework, the causal effect
of the treatment on the outcome for unit $i$ is defined as $\tau_i
\equiv Y_i(1, M_i(1)) - Y_i(0, M_i(0))$, which is typically called the
\textit{total causal effect}. Therefore, the causal mediation effect and
the total causal effect have the following relationship:
%
\begin{equation}
\tau_i  =  \delta_i(t) + \zeta_i(1-t), \label{eq:decom}
\end{equation}
where $\zeta_i(t) = Y_i(1, M_i(t)) - Y_i(0, M_i(t))$ for $t=0,1$.
This quantity $\zeta_i(t)$ is called the \textit{natural direct effect}
by Pearl  \citet{pear01} and the \textit{pure/total direct effect} by
Robins \citet{robi03}. This represents the causal effect of the
treatment on the outcome when the mediator is set to the
potential value that would occur under treatment status~$t$. In other
words, $\zeta_i(t)$ is the direct effect of the treatment when the
mediator is held constant. Equation~\eqref{eq:decom} shows an
important relationship where the total causal effect is equal to the
sum of the mediation effect under one treatment condition and the
natural direct effect under the other treatment condition. Clearly,
this equality also holds for the average total causal effect so that
$\bar\tau\equiv\E\{Y_i(1, M_i(1)) - Y_i(0, M_i(0))\} = \bar\delta(t)
+ \bar\zeta(1-t)$ for $t=0,1$ where $\bar\zeta(t)=\E(\zeta_i(t))$.

The causal mediation effects and natural direct effects differ from
the \textit{controlled direct effect} of the mediator, that is, $Y_i(t, m) -
Y_i(t, m^\prime)$ for $t=0,1$ and $m \ne m^\prime$, and that of the
treatment, that is, $Y_i(1, m) - Y_i(0, m)$ for all $m \in\mathcal{M}$
(Pearl, \citeyear{pear01}; Robins, \citep{robi03}). Unlike the mediation effects, the controlled
direct effects of the mediator are defined in terms of specific values
of the mediator, $m$ and $m^\prime$, rather than its potential values,
$M_i(1)$ and $M_i(0)$. While causal mediation analysis is used to
identify possible causal paths from $T_i$ to $Y_i$, the controlled
direct effects may be of interest, for example, if one wishes to
understand how the causal effect of $M_i$ on $Y_i$ changes as a
function of $T_i$. In other words, the former examines whether $M_i$
\textit{mediates} the causal relationship between $T_i$ and $Y_i$, whereas
the latter investigates whether $T_i$ \textit{moderates} the causal
effect of $M_i$ on $Y_i$ (Baron and Kenny, \citep{barokenn86}).

\subsection{The Main Identification Result}
\label{subsec:identification}

We now present our main identification result using the potential
outcomes framework described above. We show that under a particular
version of sequential ignorability assumption, the ACME is
nonparametrically identified. We first define our identifying
assumption:
\begin{assumption}[(Sequential ignorability)] \label{ASMIGNORABLE}
%
\begin{eqnarray}\label
{eq:YMindepT}
\{Y_i(t^\prime,m), M_i(t)\} & \indep& T_i \vert X_i = x,  \\\label{eq:YindepMgivenT}
Y_i(t^\prime,m) & \indep& M_i(t) \vert T_i = t, X_i = x
\end{eqnarray}
for $t,t^\prime= 0,1$, and all $x \in\mathcal{X}$ where it is also
assumed that $0<\Pr(T_i = t \vert X_i = x)$ and $0 < p(M_i(t) = m \vert T_i
= t, X_i = x)$ for $t = 0,1$, and all $x \in\mathcal{X}$ and $m \in
\mathcal{M}$.
\end{assumption}

Thus, the treatment is first assumed to be ignorable given
the pre-treatment covariates, and then the mediator variable is
assumed to be ignorable \textit{given} the observed value of the
treatment as well as the pre-treatment covariates. We
emphasize that, unlike the standard sequential ignorability assumption
in the literature (e.g., Robins, \citep{robi99}), the conditional
independence given in equation~\eqref{eq:YindepMgivenT} of
Assumption~\ref{ASMIGNORABLE} must hold without conditioning on the
observed values of post-treatment confounders. This issue is discussed
further below.

The following theorem presents our main identification result, showing
that under this assumption the ACME is nonparametrically identified.
\begin{theorem}[(Nonparametric identification)]\label{THIDENTIFY}
Under Assumption~\ref{ASMIGNORABLE}, the ACME and the average natural
direct effects are nonparametrically identified as follows for $t=0,1$:
%
\begin{eqnarray*}
\bar\delta(t)
& = & \int\int\E(Y_i \vert M_i = m, T_i=t, X_i = x)\\
&&\phantom{\bar\delta(t)
\;}
\{ dF_{M_i \mid T_i = 1, X_i = x}(m)\\
&&\hphantom{\bar\delta(t)
\;}{} - dF_{M_i \mid T_i = 0, X_i =
x}(m)\}\, dF_{X_i}(x),
\\
\bar\zeta(t) & = &
\int\int\{\E(Y_i | M_i = m, T_i = 1, X_i
= x)\\
&&\phantom{\bar\zeta(t)\;}
{} - \E(Y_i | M_i = m, T_i = 0, X_i = x) \}\\
&&\phantom{\bar\zeta(t) \;\,} dF_{M_i \mid T_i
= t, X_i = x}(m)\, dF_{X_i}(x),
\end{eqnarray*}
%
where $F_Z(\cdot)$ and $F_{Z\mid W}(\cdot)$ represent the
distribution function of a random variable $Z$ and the conditional
distribution function of $Z$ given $W$.
\end{theorem}

A proof is given in Appendix~\ref{app:identify}.
Theorem~\ref{THIDENTIFY} is quite general and can be easily extended
to any types of treatment regimes, for example, a continuous treatment
variable. In fact, the proof requires no change except letting $t$
and $t^\prime$ take values other than 0 and 1.
Assumption~\ref{ASMIGNORABLE} can also be somewhat relaxed by
replacing equation~\eqref{eq:YindepMgivenT} with its corresponding
mean independence assumption. However, as mentioned above, this
identification result does not hold under the standard sequential
ignorability assumption. As shown by Avin, Shpitser and
Pearl \citet{avinetal05} and also
pointed out by Robins \citet{robi03}, the nonparametric identification of
natural direct and indirect effects is not possible without an
additional assumption if equation~\eqref{eq:YindepMgivenT} holds only
after conditioning on the post-treatment confounders $Z_i$ as well as
the pre-treatment covariates $X_i$, that is, $Y_i(t^\prime, m) \indep M_i(t)
\vert T_i = t, Z_i = z, X_i = x$, for $t,t^\prime=0,1$, and all $x \in
\mathcal{X}$ and $z \in\mathcal{Z}$ where $\mathcal{Z}$ is the
support of $Z_i$. This is an important limitation since assuming the
absence of post-treatment confounders may not be credible in many
applied settings. In some cases, however, it is possible to address
the main source of confounding by conditioning on pre-treatment
variables alone (see Section~\ref{sec:emp} for an example).

\subsection{Comparison with the Existing Results in~the~Literature}
\label{subsec:literature}

Next, we compare Theorem~\ref{THIDENTIFY} with the related
identification results in the literature. First,
Pearl (\citeyear{pear01}, Theorem~2) makes the following set of assumptions in
order to identify $\bar\delta(t^\ast)$:
%
\begin{eqnarray}\label{eq:randomize}
&&p(Y(t,m) \vert X_i = x) \quad \mbox{and}\nonumber
\\[-8pt]
\\[-8pt]
&&p(M_i(t^\ast) \vert X_i = x)
\quad \mbox{are }  \mbox{identifiable}, \nonumber
\end{eqnarray}
\begin{equation}\label{eq:pearl}
Y_i(t, m) \indep M_i(t^\ast)  \vert X_i = x
\end{equation}
for all $t=0,1,$ $m \in\mathcal{M}$, and $x \in\mathcal{X}$. Under
these assumptions, Pearl arrives at the same expressions for the ACME
as the ones given in Theorem~\ref{THIDENTIFY}. Indeed, it can be
shown that Assumption~\ref{ASMIGNORABLE} implies
equations~\eqref{eq:randomize} and \eqref{eq:pearl}. While the
converse is not necessarily true, in practice, the difference is only
technical (see, e.g., Robins, \citeyear{robi03}, page~76). For example, consider a
typical situation where the treatment is randomized given the observed
pre-treatment covariates $X_i$ and researchers are interested in
identifying both $\bar\delta(1)$ and $\bar\delta(0)$. In this case,
it can be shown that Assumption~\ref{ASMIGNORABLE} is equivalent to
Pearl's assumptions.

Moreover, one practical advantage of equation~\eqref{eq:YindepMgivenT}
of Assumption~\ref{ASMIGNORABLE} is that it is easier to interpret
than equation~\eqref{eq:pearl}, which represents the independence
between the potential values of the outcome and the potential values
of the mediator. Pearl himself recognizes this difficulty, and states
``assumptions of counterfactual independencies can be meaningfully
substantiated only when cast in structural form'' (page 416). In
contrast, equation~\eqref{eq:YindepMgivenT} simply means that $M_i$ is
effectively randomly assigned given $T_i$ and $X_i$.

Second, Robins \citet{robi03} considers the identification under what he
calls a FRCISTG model, which satisfies equation~\eqref{eq:YMindepT}
as well as
%
\begin{equation}
Y_i(t,m) \indep M_i(t)   \vert  T_i = t, Z_i = z, X_i = x \label{eq:post}
\end{equation}
for $t=0,1$ where $Z_i$ is a vector of the observed values of
post-treatment variables that confound the relationship between the
mediator and outcome. The key difference between
Assumption~\ref{ASMIGNORABLE} and a FRCISTG model is that the latter
allows conditioning on $Z_i$ while the former does not.
Robins \citet{robi03} argued that this is an important practical advantage
over Pearl's conditions, in that it makes the ignorability of the
mediator more credible. In fact, not allowing for conditioning on
observed post-treatment confounders is an important limitation of
Assumption~\ref{ASMIGNORABLE}.

Under this model, Robins (\citeyear{robi03}, Theorem~2.1) shows that the
following additional assumption is sufficient to identify the ACME:
%
\begin{equation}
Y_i(1,m) - Y_i(0,m)   =   B_i, \label{eq:robins}
\end{equation}
where $B_i$ is a random variable independent of $m$. This assumption,
called the no-interaction assumption, states that the controlled
direct effect of the treatment does not depend on the value of the
mediator. In practice, this assumption can be violated in many
applications and has sometimes been regarded as ``very restrictive and
unrealistic'' (Petersen, Sinisi and van der Laan, \citep{petesinilaan06}, page 280). In contrast,
Theorem~\ref{THIDENTIFY} shows that under the sequential ignorability
assumption that does not condition on the post-treatment covariates,
the no-interaction assumption is not required for the nonparametric
identification. Therefore, there exists an important trade-off;
allowing for conditioning on observed post-treatment confounders
requires an additional assumption for the identification of the ACME.

Third, Petersen, Sinisi and van der Laan \citet{petesinilaan06} present yet another set of
identifying assumptions. In particular, they maintain
equation~\eqref{eq:YindepMgivenT} but replace
equation~\eqref{eq:YMindepT} with the following slightly weaker
condition:\looseness=1
%
\begin{eqnarray}\label{eq:YandMindepT}
Y_i(t,m) \indep T_i \vert X_i &=& x \quad  \mbox{and}\nonumber
\\[-8pt]
\\[-8pt]
 M_i(t) \indep
T_i \vert X_i &=& x\nonumber
\end{eqnarray}
for $t=0,1$ and all $m \in\mathcal{M}$. In practice, this difference
is only a technical matter because, for example, in randomized
experiments where the treatment is randomized,
equations~\eqref{eq:YMindepT} and \eqref{eq:YandMindepT} are
equivalent. However, this slight weakening of
equation~\eqref{eq:YMindepT} comes at a cost, requiring an additional
assumption for the identification of the ACME. Specifically,
Petersen, Sinisi and van der Laan \citet{petesinilaan06} assume that the magnitude of the average
direct effect does not depend on the potential values of the mediator,
that is, $\E\{Y_i(1,m) - Y_i(0,m) \vert M_i(t^\ast) = m, X_i = x\} =
\E\{Y_i(1,m) - Y_i(0,m) \vert X_i = x\}$ for all $m \in\mathcal{M}$.
Theorem~\ref{THIDENTIFY} shows that if
equation~\eqref{eq:YandMindepT} is replaced with
equation~\eqref{eq:YMindepT}, which is possible when the treatment is
randomized, then this additional assumption is unnecessary for the
nonparametric identification. In addition, this additional assumption
is somewhat difficult to interpret in practice because it entails the
mean independence relationship between the potential values of the
outcome and the potential values of the mediator.\looseness=1

Fourth, in the appendix of a recent paper, Hafeman and
VanderWeele \citet{hafevand10} show
that if the mediator is binary, the ACME can be identified with a
weaker set of assumptions than Assumption~\ref{ASMIGNORABLE}.
However, it is unclear whether this result can be generalized to cases
where the mediator is nonbinary. In contrast, the identification
result given in Theorem~\ref{THIDENTIFY} holds for any type of
mediator, whether discrete or continuous. Both identification results
hold for general treatment regimes, unlike some of the previous
results.

Finally, Rubin \citet{rubi04} suggests an alternative approach to causal
mediation analysis, which has been adopted recently by other scholars
(e.g., Egleston et al., \citep{egleetal06}; Gallop et al.,
\citep{galletal09};
Elliott, Raghunathan and
Li, \citep{ellietal10}). In this
framework, the average direct effect of the treatment is given by
$\E(Y_i(1,M_i(1)) - Y_i(0, M_i(0)) \vert M_i(1) = M_i(0))$,
representing the average treatment effect among those whose mediator
is not affected by the treatment. Unlike the average direct effect
$\bar\zeta(t)$ introduced above, this quantity is defined for a
principal stratum, which is a latent subpopulation. Within this
framework, there exists no obvious definition for the mediation effect
unless the direct effect is zero (in this case, the treatment affects
the outcome only through the mediator). Although some estimate
$\E(Y_i(1,M_i(1)) - Y_i(0, M_i(0)) \vert M_i(1) \ne M_i(0))$ and
compare it with the above average direct effect, as VanderWeele \citet{vand08a}
points out, the problem of such comparison is that two quantities are
defined for different subsets of the population. Another difficulty
of this approach is that when the mediator is continuous the
population proportion of those with $M_i(1) = M_i(0)$ can be
essentially zero. This explains why the application of this approach
has been limited to the studies with a discrete (often binary)
mediator.\looseness=1

\subsection{Implications for Linear Structural\break Equation
Model} \label{subsec:sem}

Next, we discuss the implications of Theorem~\ref{THIDENTIFY} for
LSEM, which is a popular tool among applied researchers who conduct
causal mediation analysis. In an influential article,
Baron and Kenny \citet{barokenn86} proposed a framework for mediation analysis,
which has been used by many social science methodologists; see
Mac{K}innon \citet{mack08} for a review and Imai, Keele and Tingley \citet{imaikeelting09} for a
critique of this literature. This framework is based on the
following system of linear equations:\looseness=1
%
\begin{eqnarray}\label
{eq:YgivenT}
Y_i & = & \alpha_1 + \beta_1 T_i + \epsilon_{i1}, \\\label
{eq:MgivenT}
M_i & = & \alpha_2 + \beta_2 T_i + \epsilon_{i2}, \\\label{eq:YgivenMT}
Y_i & = & \alpha_3 + \beta_3 T_i + \gamma M_i + \epsilon_{i3}.
\end{eqnarray}
Although we adhere to their original model, one may further condition
on any observed pre-treatment covariates by including them as
additional regressors in each equation. This will change none of the
results given below so long as the model includes no post-treatment
confounders.

Under this model, Baron and Kenny \citet{barokenn86} suggested that the existence of
mediation effects can be tested by separately fitting the three linear
regressions and testing the null hypotheses (1) $\beta_1 = 0$, (2)
$\beta_2 = 0$, and (3) $\gamma= 0$.
If all of these null hypotheses are rejected, they argued, then
$\beta_2 \gamma$ could be interpreted as the mediation effect.
We note that equation~\eqref{eq:YgivenT} is redundant given
equations~\eqref{eq:MgivenT} and \eqref{eq:YgivenMT}. To see this,
substitute equation~\eqref{eq:MgivenT} into
equation~\eqref{eq:YgivenMT} to obtain
%
\begin{eqnarray}\label{eq:rewrite}
Y_i  &=&  (\alpha_3 + \alpha_2 \gamma) + (\beta_3 + \beta_2 \gamma
) T_i \nonumber
\\[-8pt]
\\[-8pt]
&&{}+ (\gamma\epsilon_{i2} + \epsilon_{i3}).\nonumber
\end{eqnarray}
Thus, testing $\beta_1=0$ is unnecessary since the ACME can be
nonzero even when the average total causal effect is zero. This
happens when the mediation effect offsets the direct effect of the
treatment.

The next theorem proves that within the LSEM framework,
Baron and Kenny's interpretation is valid if
Assumption~\ref{ASMIGNORABLE} holds.
\begin{theorem}[(Identification under the LSEM)]
\label{THSEM} 
 \mbox{}  Consider the LSEM defined in
equations~\eqref{eq:YgivenT},
 \eqref{eq:MgivenT} and \eqref{eq:YgivenMT}.
Under Assumption~\ref{ASMIGNORABLE}, the ACME is identified and
given by $ \bar\delta(0) = \bar\delta(1) = \beta_2\gamma,$ where
the equality between $\bar\delta(0)$ and $\bar\delta(1)$ is also
assumed.
\end{theorem}

A proof is in Appendix~\ref{app:SEM}. The theorem implies that under
the same set of assumptions, the average natural direct effects are
identified as $\bar\zeta(0)=\bar\zeta(1)=\beta_3$, where the average
total causal effect is $\bar\tau=\beta_3 + \beta_2\gamma$. Thus,
Assumption~\ref{ASMIGNORABLE} enables the identification of the
ACME under the LSEM. Egleston et al. \citet{egleetal06} obtain a similar result
under the
assumptions of Pearl \citet{pear01} and Robins \citet{robi03}, which were reviewed
in Section~\ref{subsec:literature}.

It is important to note that under Assumption~\ref{ASMIGNORABLE}, the
standard LSEM defined in
equations~\eqref{eq:MgivenT} and \eqref{eq:YgivenMT} makes the
following no-interaction assumption about the ACME:
\begin{assumption} [(No-interaction between the Treatment
and the ACME)]
\label{asm:noint}
\[
\bar\delta(1)  =  \bar\delta(0).
\]
\end{assumption}

This assumption is equivalent to the no-interaction assumption for the
average natural direct effects,\break  $\bar\zeta(1)=\bar\zeta(0)$. Although
Assumption~\ref{asm:noint} is related to and implied by Robins'
no-interaction assumption given in equation~\eqref{eq:robins}, the key
difference is that Assumption~\ref{asm:noint} is written in terms of
the ACME rather than \textit{controlled} direct effects.

As Theorem~\ref{THIDENTIFY} suggests, Assumption~\ref{asm:noint} is
not required for the identification of the ACME under the LSEM. We
extend the outcome model given in equation~\eqref{eq:YgivenMT} to
%
\begin{equation}
Y_i  =  \alpha_3 + \beta_3 T_i + \gamma M_i + \kappa T_i M_i +
\epsilon_{i3}, \label{eq:YgivenMT2}
\end{equation}
where the interaction term between the treatment and mediating
variables is added to the outcome  regression while maintaining the
linearity in parameters. This formulation was first suggested by
Judd and Kenny \citet{juddkenn81} and more recently advocated by
Kraemer et al. (\citeyear{kraewilsfairagra02,kraekieressekupf08}) as an alternative to Barron and Kenny's
approach. Under Assumption~\ref{ASMIGNORABLE} and the
model defined by
equations~\eqref{eq:MgivenT} and \eqref{eq:YgivenMT2}, we can identify
the ACME as $\bar\delta(t) = \beta_2 (\gamma+ t \kappa)$ for $t=0,1$.
The average natural direct effects are identified as
$\bar\zeta(t)=\beta_3 + \kappa(\alpha_2 + \beta_2 t)$, and the average
total causal effect is equal to $\bar\tau=\beta_2\gamma+ \beta_3 +
\kappa(\alpha_2 + \beta_2)$. This conflicts with the proposal by
Kraemer et al. \citet{kraekieressekupf08} that the existence of mediation effects
can be established by testing either $\gamma= 0$ or $\kappa= 0$,
which is
clearly neither a necessary nor sufficient condition for $\bar\delta
(t)$ to
be zero.

The connection between the parametric and nonparametric identification
becomes clearer when both $T_i$ and $M_i$ are binary. To see this,
note that $\bar\delta(t)$ can be equivalently expressed as [dropping
the integration over $P(X_i)$ for notational simplicity]
%
\begin{eqnarray}\label{eq:discrete}
&&\bar\delta(t)  =  \sum_{m=0}^{J-1} \E(Y_i \vert M_i=m, T_i=t, X_i)
\nonumber\\
&&\hspace*{50pt}{}\cdot\{\Pr(M_i=m\vert T_i = 1, X_i)\\
&&\hspace*{60pt}{} - \Pr(M_i=m\vert T_i = 0, X_i)\}
,\nonumber
\end{eqnarray}
when $M_i$ is discrete. Furthermore, when $J=2$, this reduces to
%
\begin{eqnarray}\label{eq:binary}
\bar\delta(t)
& = & \{\Pr(M_i=1\vert T_i = 1, X_i)\nonumber\\
&&{} - \Pr(M_i=1\vert T_i =
0, X_i)\}\nonumber
\\[-8pt]
\\[-8pt]
& & {}\cdot\{\E(Y_i\vert M_i=1, T_i=t, X_i)\nonumber\\
&&\phantom{\times}{} - \E(Y_i\vert M_i=0, T_i=t,
X_i)\}.\nonumber
\end{eqnarray}
Thus, the ACME equals the product of two terms representing the
average effect of $T_i$ on $M_i$ and that of $M_i$ on $Y_i$ (holding
$T_i$ at $t$), respectively.

Finally, in the existing methodological literature Sobel \citet{sobe08}
explores the identification problem of mediation effects under the
framework of LSEM without assuming the ignorability of the mediator
(see also Albert, \citep{albe08}; Jo, \citep{jo08}). However, Sobel \citet{sobe08}
maintains, among others, the assumption that the causal effect of the
treatment is entirely through the mediator and applies the
instrumental variables technique of Angrist, Imbens and
Rubin \citet{angrimberubi96}. That
is, the natural direct effect is assumed to be zero for all units a priori, that is, $\zeta_i(t) = 0$ for all $t=0,1$ and $i$. This
assumption may be undesirable from the perspective of applied
researchers, because the existence of the natural direct effect itself
is often of interest in causal mediation analysis. See
Joffe et al. \citet{joffetal08} for an interesting application.

\section{Estimation and Inference}
\label{sec:estimation}

In this section we use our nonparametric identification result above
and propose simple parametric and nonparametric estimation strategies.

\subsection{Parametric Estimation and Inference}

Under the LSEM given by
equations~\eqref{eq:MgivenT} and \eqref{eq:YgivenMT} and
Assumption~\ref{ASMIGNORABLE}, the estimation of the ACME is
straightforward since the error terms are independent of each other.
Thus, one can follow the proposal of Baron and Kenny \citet{barokenn86} and estimate
equations~\eqref{eq:MgivenT} and \eqref{eq:YgivenMT} by fitting two
separate linear regressions. The standard error for the estimated
ACME, that is, $\hat\delta(t)=\hat\beta_2\hat\gamma$, can be calculated
either approximately using the Delta method (Sobel, \citep{sobe82}), that is,
$\var(\hat\delta(t)) \approx\beta_2^2\var(\hat{\gamma}) +
\gamma^2\var(\hat \beta_2 )$, or exactly via the variance formula of
Goodman \citet{good60}, that is, $\var(\hat\delta(t)) =
\beta_2^2\var(\hat{\gamma}) + \gamma^2\var(\hat \beta_2 ) +
\var(\hat{\gamma})\var(\hat \beta_2 )$.
For the natural direct and total effects, standard errors can be obtained
via the regressions of $Y_i$ on $T_i$ and $M_i$ [equation~\eqref
{eq:YgivenMT}] and
$Y_i$ on $T_i$ [equation~\eqref{eq:YgivenT}], respectively.

When the model contains the interaction term as in
equation~\eqref{eq:YgivenMT2} (so that Assumption~\ref{asm:noint} is
relaxed), the asymptotic variance can be computed in a similar manner.
For example, using the delta method, we have $\var(\hat\delta(t))
\approx(\gamma+ t\kappa)^2 \var(\hat\beta_2) + \beta_2^2 \{
\var(\hat\gamma) + t\var(\hat\kappa) + 2t\cov(\hat\gamma,
\hat\kappa)\}$ for $t=0,1$. Similarly,\break  $\var(\hat\zeta(t)) \approx
\var(\hat\beta_3) + (\alpha_2 + t\beta_2)^2\var(\hat\kappa)
+ 2(\alpha_2 + t\beta_2)\cov(\hat\beta_3, \hat\kappa)
+\kappa^2\{\var(\hat\alpha_2) + t\var(\hat\beta_2) + 2t\cov
\vspace*{1pt}(\hat\alpha_2,\break  \hat\beta_2)\}$.
For the average total causal effect, the variance can be obtained from
the regression of $Y_i$ on $T_i$.

\subsection{Nonparametric Estimation and Inference}
\label{subsec:nonpar}

Next, we consider a simple nonparametric estimator. Suppose that the
mediator is discrete and takes~$J$ distinct values, that is,
$\mathcal{M}=\{0,1,\ldots,J-1\}$. The case of continuous mediators is
considered further below. First, we consider the cases where we
estimate the ACME separately within each stratum defined by the
pre-treatment covariates $X_i$. One may then aggregate the resulting
stratum-specific estimates to obtain the estimated ACME. In such
situations, a nonparametric estimator can be obtained by
plugging in sample analogues for the population quantities in the expression
given in Theorem~\ref{THIDENTIFY},
%
\begin{eqnarray}\label{eq:nonpar}
&&\hat\delta(t)  =  \sum_{m=0}^{J-1} \Biggl\{\frac{\sum_{i=1}^n Y_i
\bone\{T_i = t, M_i = m\}}{\sum_{i=1}^n \bone\{T_i = t, M_i
=m\}}\nonumber
\\
&&\hspace*{50pt}{}\cdot\Biggl(\frac{1}{n_1} \sum_{i=1}^n \bone\{T_i = 1, M_i = m\}\\
&&\hspace*{68pt}{} - \frac
{1}{n_0} \sum_{i=1}^n \bone\{T_i = 0, M_i = m\} \Biggr)\Biggr\},\nonumber
\end{eqnarray}
%
where $n_t = \sum_{i=1}^n \bone\{T_i=t\}$ and $t = 0,1$. By the law of
large numbers, this estimator asymptotically converges to the true
ACME under Assumption~\ref{ASMIGNORABLE}.
The next theorem derives the asymptotic variance of the nonparametric
estimator defined in equation~\eqref{eq:nonpar} given the realized
values of the treatment variable.
\begin{theorem}[(Asymptotic variance of the nonparametric
estimator)] \label{th:var}
Suppose that
Assumption~\ref{ASMIGNORABLE} holds. Then, the variance of the
nonparametric estimator defined in equation~\eqref{eq:nonpar} is
asymptotically approximated by
\begin{eqnarray*}
\var(\hat\delta(t))
& \approx& \frac{1}{n_t}\sum_{m=0}^{J-1} \nu_{1-t,m} \biggl\{\biggl(\frac
{\nu_{1-t,m}}{\nu_{tm}}-2\biggr)\\
&&\hspace*{68pt}{}\cdot\var(Y_i \vert M_i = m, T_i = t)\\
&&\hspace*{83pt}{} + \frac
{n_t(1-\nu_{1-t,m})\mu_{tm}^2}{n_{1-t}}\biggr\} \\
& &{} - \frac{2}{n_{1-t}}\sum_{m^\prime= m +1}^{J-1} \sum_{m=0}^{J-2}
\nu_{1-t,m}\nu_{1-t,m^\prime}\mu_{tm}\mu_{tm^\prime}\\
&&{} + \frac
{1}{n_t}\var(Y_i \vert T_i = t)
\end{eqnarray*}
for $t=0,1$ where $\nu_{tm} \equiv\Pr(M_i = m \vert T_i = t)$
and $\mu_{tm} \equiv\E(Y_i \vert M_i = m, T_i = t)$.
\end{theorem}

A proof is based on a tedious but simple application of the Delta
method and thus is omitted. This asymptotic variance can be
consistently estimated by replacing unknown population quantities with
their corresponding sample counterparts. The estimated overall
variance can be obtained by aggregating the estimated within-strata
variances according to the sample size in each stratum.

\begin{table*}[b]
\tabcolsep=0pt
\caption{Finite-sample performance of the proposed estimators and their
variance estimators. The table presents the results of a Monte Carlo experiment
with varying sample sizes and fifty thousand iterations. The upper half
of the table
represents the results for $\hat\delta(0)$ and the bottom half $\hat
\delta(1)$.
The columns represent (from left to right) the following: sample sizes,
estimated biases,
root mean squared errors (RMSE) and the coverage
probabilities of the 95\% confidence intervals of the nonparametric estimators,
and the same set of quantities for the parametric estimators. The true
values of
$\bar\delta(0)$ and $\bar\delta(1)$ are $0.675$ and $4.03$,
respectively. The results indicate that nonparametric estimators have
smaller bias than the parametric estimator though its variance is much
larger. The confidence intervals converge to the nominal coverage as
the sample size increases. The convergence occurs much more quickly for
the parametric estimator}
\label{tab:varsim}
\begin{tabular*}{\textwidth}{@{\extracolsep{\fill}}ld{3.0}d{2.3}ccd{2.3}cc@{}}
\hline
& & \multicolumn{3}{c}{\textbf{Nonparametric estimator}} &  \multicolumn
{3}{c}{\textbf{Parametric estimator}} \\
\ccline{3-5,6-8}
& \multicolumn{1}{c}{\textbf{Sample size}} & \multicolumn{1}{c}{\textbf{Bias}} & \textbf{RMSE} & \textbf{95\% CI coverage} &  \multicolumn{1}{c}{\textbf{Bias}}
& \textbf{RMSE} & \textbf{95\%
CI coverage} \\
\hline
$\hat\delta(0)$ & 50 &  0.002 & 1.034 & 0.824 &   0.096 &
0.965 & 0.919 \\
& 100 &  0.006 & 0.683 & 0.871 &   0.044 & 0.566 & 0.933 \\
& 500 & -0.002 & 0.292 & 0.922 &   0.006 & 0.229 & 0.947 \\
[3pt]
$\hat\delta(1)$ & 50 &  0.010 & 2.082 & 0.886  & -0.010 & 1.840
& 0.934 \\
& 100 &  0.005 & 1.462 & 0.912  &  0.003 & 1.290 & 0.944 \\
& 500 &  0.001 & 0.643 & 0.939  &  0.001 & 0.570 & 0.955 \\
\hline
\end{tabular*}
\end{table*}

The second and perhaps more general strategy is to use nonparametric
regressions to model $\mu_{tm}(x) \equiv\E(Y_i \vert T_i = t, M_i = m,
X_i = x)$ and $\nu_{tm}(x)\equiv\Pr(M_i = m \vert T_i = t, X_i =
x)$, and then employ the following estimator:
%
\begin{eqnarray}\label{eq:nonparX}
&&\hat\delta(t)  =  \frac{1}{n}\Biggl \{\sum_{i=1}^n \sum_{m=0}^{J-1}
\hat\mu_{tm}(X_i)\nonumber
\\[-8pt]
\\[-8pt]
 &&\hspace*{78pt}{}\cdot\bigl(\hat\nu_{1m}(X_i) - \hat\nu_{0m}(X_i)\bigr)
\Biggr\} \nonumber
\end{eqnarray}
for $t=0,1$. This estimator is also asymptotically consistent for the
ACME under Assumption~\ref{ASMIGNORABLE}
if $\hat\mu_{tm}(x)$ and $\hat\nu_{tm}(x)$ are consistent for
$\mu_{tm}(x)$ and $\nu_{tm}(x)$, respectively.
Unfortunately, in general, there is no simple expression for the
asymptotic variance of this estimator. Thus, one may use a
nonparametric bootstrap [or a parametric bootstrap based on the
asymptotic distribution of $\hat\mu_{tm}(x)$ and $\hat\nu_{tm}(x)$] to
compute uncertainty estimates.

Finally, when the mediator is not discrete, we may nonparametrically
model $\mu_{tm}(x) \equiv\E(Y_i \vert T_i = t, M_i = m, X_i = x)$ and
$\psi_t(x) = p(M_i \vert T_i = t, X_i = x)$. Then, one can use the
following estimator:
%
\begin{equation}\hspace*{15pt}
\hat\delta(t)  =  \frac{1}{nK} \sum_{i=1}^n
\sum_{k=1}^K \bigl\{\hat\mu_{t\tilde{m}_{1i}^{(k)}}(X_i) -
\hat\mu_{t\tilde{m}_{0i}^{(k)}}(X_i)\bigr\},
\end{equation}
where $\tilde{m}_{ti}^{(k)}$ is the $k$th Monte Carlo draw of the
mediator~$M_i$ from its predicted distribution based on the fitted
model $\hat\psi_t(X_i)$.

These estimation strategies are quite general in that they can be
applied to a wide range of statistical models.
Imai, Keele and Tingley \citet{imaikeelting09} demonstrate the generality of these
strategies by applying them to common parametric and nonparametric
regression techniques often used by applied researchers. By doing so,
they resolve some confusions held by social science methodologists,
for example, how to estimate mediation effects when the outcome and/or
the mediator is binary. Furthermore, the proposed general estimation
strategies enable Imai et al. \citet{imaietal10} to develop an easy-to-use R
package, \texttt{mediation}, that implements these methods and
demonstrate its use with an empirical example.

\subsection{A Simulation Study}
\label{subsec:sims}

Next, we conduct a small-scale Monte Carlo experiment in order to
investigate the finite-sample performance of the estimators defined in
equations~\eqref{eq:nonpar} and \eqref{eq:nonparX} as well as the
proposed variance estimator given in Theorem~\ref{th:var}. We use a
population model where the potential outcomes and mediators are given
by $Y_i(t,m)=\exp(Y_i^\ast(t,m))$, $M_i(t)=\bone\{M_i^\ast(t)\geq
0.5\}$ and $Y_i^\ast(t,m)$, $M_i^\ast(t)$ are jointly normally
distributed. The population parameters are set to the following
values: $\E(Y_i^\ast(1,1))=2$; $\E(Y_i^\ast(1,0))=0$;
$\E(Y_i^\ast(0,1))=1$; $\E(Y_i^\ast(0,0))=0.5$; $\E(M_i^\ast(1))=1$;
$\E(M_i^\ast(0))=0$; $\var(Y_i^\ast(t,m))=\var(M_i^\ast(t))=1$ for
$t\in\{0,1\}$ and $m\in\{0,1\}$; $\operatorname{Corr}(Y_i^\ast(t,m),
Y_i^\ast(t^\prime,m^\prime))=0.5$ for $t,t^\prime\in\{0,1\}$ and $m,
m^\prime\in\{0,1\}$; $\operatorname{Corr}(Y_i^\ast(t,m),M_i^\ast(t^{\prime}))=0$ for $t\in\{0,1\}$ and
$m\in\{0,1\}$; and $\operatorname{Corr}(M_i^\ast(1),\break M_i^\ast(0)) = 0.3$.

Under this setup, Assumption~\ref{ASMIGNORABLE} is satisfied. Thus,
we can consistently estimate the ACME by applying the nonparametric
estimator given in equation~\eqref{eq:nonpar}. Also, note that this
data generating process implies the following parametric regression
models for the observed data:
%
\begin{eqnarray} \label{eq:Mprobit}
 \hspace*{27pt} \Pr(M_i = 1 \vert T_i) & = & \Phi(\alpha_2 + \beta_2T_i), \\\label{eq:Ylognorm} \hspace*{27pt}
Y_i \vert T_i, M_i & \sim& \operatorname{lognormal}(\alpha_3 + \beta_3T_i
+ \gamma M_i \nonumber
\\[-8pt]
\\[-8pt]  \hspace*{27pt}&& \phantom{\operatorname{lognormal}(\alpha_3 +}{}+ \kappa T_iM_i, \sigma^2_3),\nonumber
\end{eqnarray}
where $(\alpha_2, \beta_2, \alpha_3, \beta_3, \gamma, \kappa,
\sigma_3^2) = (-0.5, 1, 0.5, -0.5,\break  0.5, 1.5, 1)$ and $\Phi(\cdot)$ is
the standard normal distribution function. We can then obtain the
parametric maximum likelihood estimate of the ACME by fitting these
two models via standard procedures and estimating the following
expression based on Theorem~\ref{THIDENTIFY} [see
equation~\eqref{eq:binary}]:
%
\begin{eqnarray}\label{eq:simtrue}
\bar\delta(t)  &=&  \{\exp(\alpha_3 + \beta_3t + \gamma+
\kappa t + \sigma^2_3/2)\nonumber\\
 &&\hspace*{33pt}{}- \exp(\alpha_3 + \beta_3t
+ \sigma^2_3/2)\}
\\
&&{}\cdot\{\Phi(\alpha_2 + \beta_2) -
\Phi(\alpha_2)\}\nonumber
\end{eqnarray}
for $t=0,1$.

We compare the performances of these two estimators via Monte Carlo
simulations. Specifically, we set the sample size $n$ to 50, 100 and
500 where half of the sample receives the treatment and the other half
is assigned to the control group, that is, $n_1=n_0=n/2$. Using
equation~\eqref{eq:simtrue}, the true values of the ACME are given by
$\bar\delta(0) = 0.675$ and $\bar\delta(1) = 4.03$.

Table~\ref{tab:varsim} reports the results of the experiments based on
fifty thousand iterations. The performance of the estimators turns
out to be quite good in this particular setting. Even with sample
size as small as 50, estimated biases are essentially zero for the
nonparametric estimates. The parametric estimators are\break slightly more
biased for the small sample sizes, but they converge to the true
values by the time the sample size reaches 500. As expected, the
variance is larger for the nonparametric estimator than the parametric
estimator. The 95\% confidence intervals converge to the nominal
coverage as the sample size increases. The convergence occurs much
more quickly for the parametric estimator. (Although not reported in
the table, we confirmed that for both estimators the coverage
probabilities fully converged to their nominal values by the time the
sample size reached 5000.)

\section{Sensitivity Analysis}
\label{sec:sens}

Although the ACME is nonparametrically identified under
Assumption~\ref{ASMIGNORABLE}, this assumption, like other existing
identifying assumptions, may be too strong in many applied settings.
Consider randomized experiments where the treatment is randomized but
the mediator is not. Causal mediation analysis is most frequently
applied to such experiments. In this case,
equation~\eqref{eq:YMindepT} of Assumption~\ref{ASMIGNORABLE} is
satisfied but equation~\eqref{eq:YindepMgivenT} may not hold for two
reasons. First, there may exist unmeasured pre-treatment covariates
that confound the relationship between the mediator and the outcome.
Second, there may exist observed or unobserved post-treatment
confounders. These possibilities, along with other obstacles
encountered in applied research, have led some scholars to warn
against the abuse of mediation analyses
(e.g., Green, Ha and
Bullock, \citep{greehabull10}). Indeed, as we formally show below,
the data generating process contains no information about the
credibility of the sequential ignorability assumption.

To address this problem, we develop a method to
assess the sensitivity of an estimated ACME to unmeasured
pre-treatment confounding (The proposed sensitivity analysis, however,
does not address the possible existence of post-treatment
confounders). The method is based on the standard LSEM framework
described in Section~\ref{subsec:sem} and can be easily used by
applied researchers to examine the robustness of their empirical
findings. We derive the maximum departure from
equation~\eqref{eq:YindepMgivenT} that is allowed while maintaining
their original conclusion about the direction of the ACME
(see Imai and Yamamoto, \citep{imaiyama10}). For notational simplicity, we do not
explicitly condition on the pre-treatment covariates $X_i$. However,
the same analysis can be conducted by including them as additional
covariates in each regression.

\subsection{Parametric Sensitivity Analysis Based on the Residual Correlation}

The proof of Theorem~\ref{THSEM} implies that if
equation~\eqref{eq:YMindepT} holds, $\epsilon_{i2} \indep T_i$ and
$\epsilon_{i3} \indep T_i$ hold but $\epsilon_{i2} \indep
\epsilon_{i3}$ does not unless equation~\eqref{eq:YindepMgivenT} also
holds. Thus, one way to assess the sensitivity of one's conclusions
to the violation of equation~\eqref{eq:YindepMgivenT} is to use the
following sensitivity parameter:
%
\begin{equation}
\rho \equiv \operatorname{Corr}(\epsilon_{i2}, \epsilon_{i3}), \label{eq:corr}
\end{equation}
where $-1 < \rho< 1$. In Appendix~\ref{app:rho} we show that
Assumption~\ref{ASMIGNORABLE} implies $\rho=0$. (Of course, the
contrapositive of this statement is also true; $\rho\ne0$ implies
the violation of Assumption~\ref{ASMIGNORABLE}). A nonzero
correlation parameter can be interpreted as the existence of omitted
variables that are related to both the observed value of the mediator
$M_i$ and the potential outcomes $Y_i$ even after conditioning on the
treatment variable $T_i$ (and the observed covariates $X_i$). Note
that these omitted variables must causally precede $T_i$. Then, we
vary the value of $\rho$ and compute the corresponding estimate of the
ACME. In a quite different context, Roy, Hogan and Marcus \citet{royetal08} take
this general strategy of computing a quantity of interest at various
values of an unidentifiable sensitivity parameter.

The next theorem shows that if the treatment is randomized, the
ACME is identified given a particular value of $\rho$.
\begin{theorem}[(Identification with a given error
correlation)] \label{THMIDENTIFY1} 
Consider the
LSEM defined in
equations \eqref{eq:YgivenT}, \eqref{eq:MgivenT} and \eqref{eq:YgivenMT}.
Suppose that equation~\eqref{eq:YMindepT} holds and the correlation
between $\epsilon_{i2}$ and $\epsilon_{i3}$, that is, $\rho$, is given.
If we further assume $-1 < \rho  < 1$, then the ACME is identified
and given by
\[
\bar\delta(0)  =  \bar\delta(1)  =  \frac{\beta_2\sigma
_1}{\sigma_2} \bigl\{\tilde\rho-\rho\sqrt{(1-\tilde\rho^2)/(1-\rho
^2)} \bigr\},
\]
where $\sigma_j^2 \equiv\var(\epsilon_{ij})$ for $j=1,2$ and
$\tilde\rho\equiv\operatorname{Corr}(\epsilon_{i1},\break \epsilon_{i2})$.
\end{theorem}

A proof is in Appendix~\ref{app:identify1}. We offer several remarks
about Theorem~\ref{THMIDENTIFY1}. First, the unbiased estimates of
$(\alpha_1, \alpha_2, \beta_1, \beta_2)$ can be obtained by fitting
the equation-by-equation least squares of
equations \eqref{eq:YgivenT} and \eqref{eq:MgivenT}. Given these
estimates, the covariance matrix of $(\epsilon_{i1}, \epsilon_{i2})$,
whose elements are $(\sigma_1^2, \sigma_2^2,\break
\tilde\rho\sigma_1\sigma_2)$, can be consistently estimated by
computing the sample covariance matrix of the residuals, that is,
$\hat\epsilon_{i1} = Y_i - \hat\alpha_1 - \hat\beta_1 T_i$ and
$\hat\epsilon_{i2} = M_i - \hat\alpha_2 - \hat\beta_2T_i$.

Second, the partial derivative of the ACME with respect to $\rho$
implies that the ACME is either monotonically increasing or decreasing
in $\rho$, depending on the sign of $\beta_2$. The ACME is also
symmetric about $(\rho, \bar\delta(t))=(0, \beta_2
\tilde\rho\sigma_1/\sigma_2)$.

Third, the ACME is zero if and only if $\rho$ equals $\tilde\rho$.
This implies that researchers can easily check the robustness of their
conclusion obtained under the sequential ignorability assumption via
correlation between $\epsilon_{i1}$ and $\epsilon_{i2}$. For example,
if $\hat{\delta}(t) = \hat{\beta}_2\hat{\gamma}$ is negative, the true
ACME is also guaranteed to be negative if $\rho< \tilde\rho$ holds.

Fourth, the expression of the ACME given in
Theorem~\ref{THMIDENTIFY1} is cumbersome to use when computing the
standard errors. A more straightforward and general approach is to
apply the iterative feasible generalized least square algorithm of the
seemingly unrelated regression (Zellner, \citep{zell62}), and use the associated
asymptotic variance formula. This strategy will also work when there
is an interaction term between the treatment and mediating variables
as in equation~\eqref{eq:YgivenMT2} and/or when there are observed
pre-treatment covariates $X_i$.

Finally, Theorem~\ref{THMIDENTIFY1} implies the following corollary,
which shows that under the LSEM the data generating process is not
informative at all about either the sensitivity parameter $\rho$ or
the ACME without equation~\eqref{eq:YindepMgivenT}. This result
highlights the
difficulty of causal mediation analysis and the importance of
sensitivity analysis even in the parametric modeling setting.
\begin{corollary}[(Bounds on the sensitivity parameter)]
\label{cor:identify1} 
Consider the LSEM defined in
equations~\eqref{eq:YgivenT}, \eqref{eq:MgivenT} and \eqref{eq:YgivenMT}.
Suppose that equation~\eqref{eq:YMindepT} holds but
equation~\eqref{eq:YindepMgivenT} may not. Then, the sharp, that is,
best possible, bounds on the sensitivity parameter $\rho$ and ACME
are given by $(-1,1)$ and $(-\infty, \infty)$, respectively.
\end{corollary}

The first statement of the corollary follows directly from the proof
of Theorem~\ref{THMIDENTIFY1}, while the second statement can be
proved by taking a limit of $\delta(t)$ as $\rho$ tends to $-1$ or
$1$.

\subsection{Parametric Sensitivity Analysis Based on the Coefficients
of Determination}

The sensitivity parameter $\rho$ can be given an alternative
definition which allows it to be interpreted as the magnitude of an
unobserved confounder. This alternative version of $\rho$ is based on
the following decomposition of the error terms in
equations~\eqref{eq:MgivenT} and \eqref{eq:YgivenMT}:
\begin{eqnarray*}
\epsilon_{ij} & = & \lambda_j U_i + \epsilon_{ij}^\prime
\end{eqnarray*}
for $j=2,3$, where $U_i$ is an unobserved confounder and the
sequential ignorability is assumed given $U_i$ and $T_i$. Again, note
that $U_i$ has to be a pre-treatment variable so that the resulting
estimates can be given a causal interpretation. In addition, we
assume that $\epsilon_{ij}^\prime\indep U_i$ for $j=2,3$. We can
then express the influence of the unobserved pre-treatment confounder
using the following coefficients of determination:
\[
R_M^{2*}  \equiv 1-\frac{\var(\epsilon_{i2}^\prime)}{\var
(\epsilon_{i2})} \]
and
\[
R_Y^{2*}  \equiv 1-\frac{\var(\epsilon_{i3}^\prime)}{\var
(\epsilon_{i3})},
\]
which represent the proportion of previously unexplained variance
(either in the mediator or in the outcome) that is explained by the
unobserved confounder (see Imbens, \citep{imbe03}).

\begin{table*}[b]
\tablewidth=406pt
\caption{Parametric and nonparametric estimates of the ACME under
sequential ignorability in the media framing experiment. Each
cell of the table represents an estimated average causal effect
and its 95\% confidence interval. The outcome is the subjects'
tolerance level
for the free speech rights of the Ku Klux Klan, and the treatments are the
public order frame ($T_i = 1$) and the free speech frame ($T_i=0$).
The second column of the table shows the results of the parametric LSEM
approach, while
the third column of the table presents those of the nonparametric estimator.
The lower part of the table shows the results of parametric mediation analysis
under the no-interaction assumption [$\hat\delta(1) = \hat\delta
(0)$], while\vspace*{1pt}
the upper part presents the findings without this assumption, thereby showing
the estimated average mediation effects under the treatment and the control,
that is, $\hat\delta(1)$ and $\hat\delta(0)$}
\label{tab:nelson1}
\begin{tabular*}{406pt}{@{\extracolsep{\fill}}lcc@{}}
\hline
& \multicolumn{1}{c}{\textbf{Parametric}} & \multicolumn{1}{c@{}}{\textbf{Nonparametric}}
\\
\hline
Average mediation effects\\
 \quad  Free speech frame $\hat\delta(0)$ & $-0.566$ & $-0.596$ \\
&\multicolumn{1}{c}{$[-1.081$, $-0.050]$}&\multicolumn{1}{c@{}}{$[-1.168$, $-0.024]$}\\[3pt]
\quad   Public order frame $\hat\delta(1)$& $-0.451$ & $-0.374$ \\
&\multicolumn{1}{c}{$[-0.871$, $-0.031]$}&\multicolumn{1}{c@{}}{$[-0.823$,  $0.074]$} \\[3pt]
\quad Average total effect $\hat\tau$ & $-0.540$ & $-0.540$ \\
&\multicolumn{1}{c}{$[-1.207$, $ 0.127]$}&\multicolumn{1}{c@{}}{$[-1.206$,  $0.126]$}\\[6pt]
With the no-interaction assumption \\
\quad Average mediation effect & $-0.510$ & \\
 \multicolumn{1}{c}{$\hat\delta(0)=\hat\delta(1)$} & \multicolumn{1}{c}{$[-0.969$, $-0.051]$} & \\[3pt]
\quad Average total effect $\hat\tau$ & $-0.540$\\
& \multicolumn{1}{c}{$[-1.206$,  $0.126]$} & \\
\hline
\end{tabular*}
\end{table*}

Another interpretation is based on the proportion of original variance
that is explained by the unobserved confounder. In this case, we use
the following sensitivity parameters:
\[
\widetilde{R}_M^2  \equiv \frac{\var(\epsilon_{i2})-\var
(\epsilon_{i2}^\prime)}{\var(M_i)} =  (1-R^2_M)R_M^{2*}
\]
and
\[
\widetilde{R}_Y^2  \equiv \frac{\var(\epsilon_{i3})-\var
(\epsilon_{i3}^\prime)}{\var(Y_i)}  =  (1-R^2_Y)R_Y^{2*},
\]
where $R_M^2$ and $R_Y^2$ represent the coefficients of determination
from the two regressions given in
equations~\eqref{eq:MgivenT} and \eqref{eq:YgivenMT}. Note that
unlike\vspace*{1.5pt} $R_M^{2*}$ and $R_Y^{2*}$ (as well as $\rho$ given in
Corollary~\ref{cor:identify1}), $\widetilde{R}_M^2$ and
$\widetilde{R}_Y^2$ are bounded from above by
$\var(\epsilon_{i2})/\var(M_i)$ and $\var(\epsilon_{i3})/\var(Y_i)$,
respectively.

In either case, it is straightforward to show that the following
relationship between $\rho$ and these\vspace*{1pt} parameters holds, that is, $\rho^2
= R_M^{2*}R_Y^{2*} =
\widetilde{R}_M^2\widetilde{R}_Y^2/\{(1-R^2_M)(1-R^2_Y)\}$ or,
equivalently,
\[
\rho =  \sgn(\lambda_2\lambda_3)R_M^*R_Y^*  =  \frac{\sgn
(\lambda_2\lambda_3)\widetilde{R}_M\widetilde{R}_Y}{\sqrt
{(1-R^2_M)(1-R^2_Y)}},
\]
where $R_M^*, R_Y^*, \widetilde{R}_M$ and $\widetilde{R}_Y$ are in $[0,1]$.
Thus, in this framework, researchers can specify the values of
$(R_M^{2*},R_Y^{2*})$ or $(\widetilde{R}_M^2,\widetilde{R}_Y^2)$ as
well as the sign of $\lambda_2 \lambda_3$ in order to determine values
of $\rho$ and estimate the ACME based on these values of $\rho$. Then,
the analyst can examine variation in the estimated ACME with respect to
change in these parameters.

\subsection{Extensions to Nonlinear and Nonparametric~Models}

The proposed sensitivity analysis above is developed within the
framework of the LSEM, but some extensions are possible. For example,
Imai, Keele and Tingley \citet{imaikeelting09} show how to conduct sensitivity analysis
with probit models when the mediator and/or the outcome are discrete.
In Appendix~\ref{app:nonparasens}, while it is substantially more
difficult to conduct such an analysis in the nonparametric setting, we
consider sensitivity analysis for the nonparametric plug-in estimator
introduced in Section~\ref{subsec:nonpar} (see also VanderWeele, \citep{vand10}
for an alternative approach).

\section{Empirical Application}
\label{sec:emp}

In this section we apply our proposed methods to the influential
randomized experiment from political psychology we described in
Section~\ref{sec:example}.

\subsection{Analysis under Sequential Ignorability}
\label{subsec:anl}

In the original analysis, Nelson, Clawson and
Oxley \citet{nelsclawoxle97} used a LSEM similar
to the one discussed in Section~\ref{subsec:sem} and found that
subjects who viewed the Klan story with the free speech frame were
significantly more tolerant of the Klan than those who saw the story
with the public order frame. The researchers also found evidence
supporting their main hypothesis that subjects' general attitudes
mediated the causal effect of the news story frame on tolerance for
the Klan. In the analysis that follows, we only analyze the public
order mediator, for which the researchers found a significant mediation
effect.

As we showed in Section~\ref{subsec:sem}, the original results can be given
a causal interpretation under sequential ignorability, that is,
Assumption~\ref{ASMIGNORABLE}.
Here, we first make this assumption and estimate
causal effects based on our theoretical results.
Table~\ref{tab:nelson1} presents the findings. The second and
third columns of the table show the estimated ACME and average total effect
based on the LSEM and the nonparametric estimator, respectively.
The 95\% asymptotic
confidence intervals are constructed using the Delta method. For most
of the estimates, the 95\% confidence intervals do not contain zero,
mirroring the finding from the original study that general attitudes
about public order mediated the effect of the media frame.

As shown in Section~\ref{subsec:sem}, we can relax the no-interaction
assumption (Assumption~\ref{asm:noint}) that is implicit in the
LSEM of Baron and Kenny \citet{barokenn86}.
The first and second rows of the table present estimates from the
parametric and nonparametric analysis without this
assumption. These results show that the estimated
ACME under the free speech condition [$\hat\delta(0)$] is
larger than the effect under the public order condition
[$\hat\delta(1)$] for both the parametric and nonparametric
estimators. In fact, the 95\% confidence interval for the
nonparametric estimate of $\bar\delta(1)$ includes zero. However, we
fail to reject the null hypothesis of $\bar\delta(0)=\bar\delta(1)$
under the parametric analysis, with a $p$-value of $0.238$.

Based on this finding, the no-interaction assumption could be regarded
as appropriate. The last two rows in Table~\ref{tab:nelson1} contain
the analysis
based on the parametric estimator under this assumption. As expected,
the estimated ACME is between the previous two estimates,
and the 95\% confidence interval does not contain zero. Finally, the estimated
average total effect is identical to that without Assumption~\ref{asm:noint}.
This makes sense since the no-interaction assumption only restricts the way
the treatment effect is transmitted to the outcome and thus does not
affect the
estimate of the overall treatment effect.

\subsection{Sensitivity Analysis}
\label{subsec:emp-sens}

The estimates in Section~\ref{subsec:anl} are identified if the
sequential ignorability assumption holds. However, since the original
researchers randomized news stories but subjects' attitudes were
merely observed, it is unlikely this assumption holds. As we
discussed in Section~\ref{sec:example}, one particular concern is that
subjects' pre-existing ideology affects both their attitudes toward
public order issues and their tolerance for the Klan within each
treatment condition. Thus, we next ask how sensitive these estimates
are to violations of this assumption using the methods proposed in
Section~\ref{sec:sens}. We consider political ideology to be a
possible unobserved pre-treatment confounder. We also maintain
Assumption~\ref{asm:noint}.

\begin{figure}[b]

\includegraphics{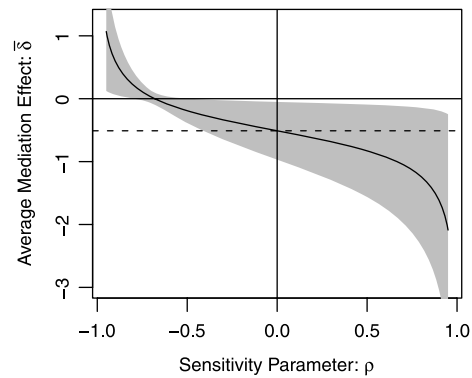}

\caption{Sensitivity analysis for the media framing experiment. The
figure presents the results of the sensitivity analysis described in
Section~\protect\ref{sec:sens}. The solid line represents the estimated
ACME for the attitude mediator for differing values of the
sensitivity parameter $\rho$, which is defined in
equation~\protect\eqref{eq:corr}. The gray region represents the 95\%
confidence interval based on the Delta method. The horizontal
dashed line is drawn at the point estimate of $\bar\delta$ under
Assumption~\protect\ref{ASMIGNORABLE}.} \label{psensfig}
\end{figure}

Figure~\ref{psensfig} presents the results for the sensitivity
analysis based on the residual correlation. We plot the estimated
ACME of the attitude mediator against differing values of the
sensitivity parameter $\rho$, which is equal to the correlation
between the two error terms of
equations~(\ref{eq:MgivenT1}) and (\ref{eq:YgivenMT1}) for each. The
analysis indicates that the original conclusion about the direction of
the ACME under Assumption~\ref{ASMIGNORABLE} (represented by the
dashed horizontal line) would be maintained unless $\rho$ is less than
$-0.68$. This implies that the conclusion is plausible given even
fairly large departures from the ignorability of the mediator. This
result holds even after we take into account the sampling variability,
as the confidence interval covers the value of zero only when $-0.79 <
\rho< -0.49$. Thus, the original finding about the negative ACME is
relatively robust to the violation of
equation~\eqref{eq:YindepMgivenT} of Assumption~\ref{ASMIGNORABLE}
under the LSEM.

\clearpage
\begin{sidewaysfigure}
\vspace*{280pt}
\includegraphics{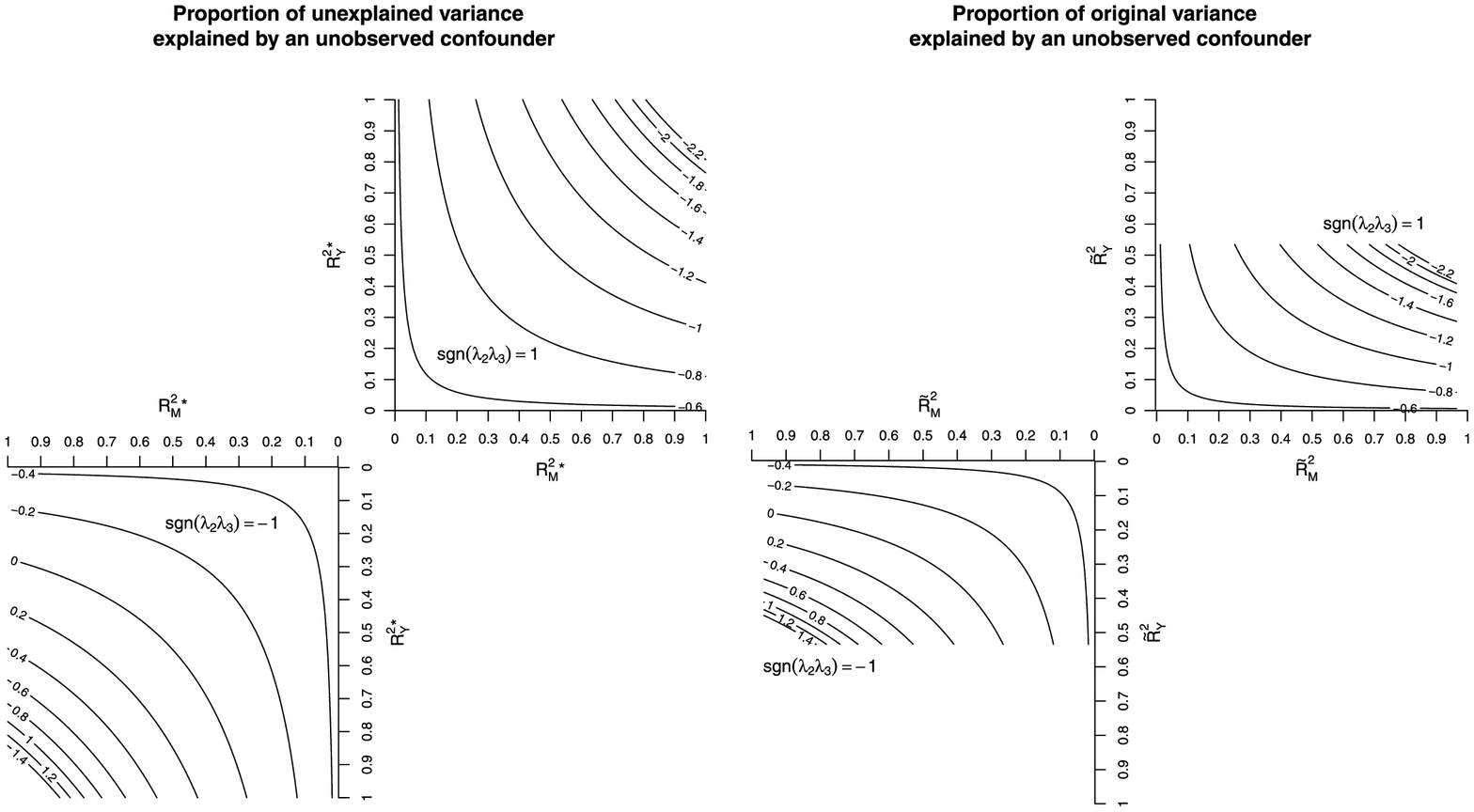}

\caption{An alternative interpretation of the sensitivity analysis.
The plot presents the results of the sensitivity analysis described
in Section~\protect\ref{sec:sens}. Each plot contains various\vspace*{1pt} mediation
effects under an unobserved pre-treatment confounder of various
magnitudes. The left plot contains the contours for $R^{2*}_M$ and
$R^{2*}_Y$ which represent the proportion of unexplained variance
that is explained by the unobserved confounder for the mediator and
outcome, respectively. The right plot contains the contours for
$\tilde{R}^2_M$ and $\tilde{R}^2_Y$ which represent the proportion
of the variance explained by the unobserved pre-treatment
confounder. Each line represents the estimated ACME under proposed
values of either $(R^{*2}_M, R^{2*}_Y)$ or $(\tilde{R}^2_M,
\tilde{R}^2_Y)$. The term $\sgn(\lambda_2\lambda_3)$ represents the
sign on the product of the coefficients of the unobserved
confounder.} \label{psensfig2}
\end{sidewaysfigure}
\clearpage

Next, we present the same sensitivity analysis using the alternative
interpretation of $\rho$ which is based on two coefficients of
determination as defined in Section~\ref{sec:sens}; (1) the proportion
of unexplained variance that is explained by an unobserved
pre-treatment confounder ($R^{2*}_M$ and $R^{2*}_Y$) and (2) the
proportion of the original variance explained by the same unobserved
confounder ($\tilde{R}^2_M$ and $\tilde{R}^2_Y$).
Figure~\ref{psensfig2} shows two plots based on the types of
coefficients of determination. The lower left quadrant of each plot
in the figure represents the case where the product of the
coefficients for the unobserved confounder is negative, while the upper
right quadrant represents the case where the product is positive.

For example, this product will be positive if the unobserved
pre-treatment confounder represents subjects' political ideology,
since conservatism is likely to be positively correlated with both
public order importance and tolerance for the Klan. Under this
scenario, the original conclusion about the direction of the ACME is
perfectly robust to the violation of sequential ignorability, because
the estimated ACME is always negative in the upper right quadrant
of each plot. On the other hand, the result is less robust to the
existence of an unobserved confounder that has opposite effects
on the mediator and outcome. However, even for this alternative
situation, the ACME is still guaranteed to be negative as long as the
unobserved confounder explains less than 27.7\% of the variance in the
mediator or outcome that is left unexplained by the treatment alone,
no matter how large the corresponding portion of the variance in the
other variable may be. Similarly, the direction of the original
estimate is maintained if the unobserved confounder explains less than
26.7\% (14.7\%) of the original variance in the mediator (outcome),
regardless of the degree of confounding for the outcome (mediator).

\section{Concluding Remarks}
\label{sec:conclusion}

In this paper we study identification, inference and sensitivity
analysis for causal mediation effects. Causal mediation analysis is
routinely conducted in various disciplines, and our paper contributes
to this fast-growing methodological literature in several ways.
First, we provide a new identification condition for the ACME, which
is relatively easy to interpret in substantive terms and also weaker
than existing results in some situations. Second, we prove that the
estimates based on the standard LSEM can be given valid causal
interpretations under our proposed framework. This provides a basis
for formally analyzing the validity of empirical studies using the
LSEM framework. Third, we propose simple nonparametric estimation
strategies for the ACME. This allows researchers to avoid the
stronger functional form assumptions required in the standard LSEM.
Finally, we offer a parametric sensitivity analysis that can be easily
used by applied researchers in order to assess the sensitivity of
estimates to the violation of this assumption. We view sensitivity
analysis as an essential part of causal mediation analysis because the
assumptions required for identifying causal mediation effects are
unverifiable and often are not justified in applied settings.

At this point, it is worth briefly considering the progression of
mediation research from its roots in the empirical psychology
literature to the present. In their seminal paper,
Baron and Kenny \citet{barokenn86} supplied applied researchers with a simple method
for mediation analysis. This method has quickly gained widespread
acceptance in a number of applied fields. While psychologists
extended this LSEM framework in a number of ways, little attention was
paid to the conditions under which their popular estimator can be
given a causal interpretation. Indeed, the formal definition of the
concept of causal mediation had to await the later works by
epidemiologists and statisticians (Robins and Greenland, \citep{robigree92}; Pearl, \citep{pear01}; Robins, \citep{robi03}).
The progress made on the identification of causal
mediation effects by these authors has led to the recent development
of alternative and more general estimation strategies
(e.g., Imai, Keele and Tingley, \citep{imaikeelting09}; VanderWeele, \citep{vand09}). In this paper we show
that under a set of assumptions this popular product of coefficients
estimator can be given a causal interpretation. Thus, over twenty
years later, the work of Baron and Kenny has come full circle.\looseness=1

Despite its natural appeal to applied scientists, statisticians often
find the concept of causal mediation mysterious
(e.g., Rubin, \citep{rubi04}). Part of this skepticism seems to stem from
the concept's inherent dependence on background scientific
theory;\break
whether a variable qualifies as a mediator in a given empirical study
relies crucially on the investigator's belief in the theory being
considered. For example, in the social science application introduced
in Section~\ref{sec:example}, the original authors test whether the
effect of a media framing on citizens' opinion about the Klan rally is
mediated by a change in attitudes about general issues. Such a setup
might make no sense to another political psychologist who hypothesizes
that the change in citizens' opinion about the Klan rally prompts
shifts in their attitudes about more general underlying issues. The
H1N1 flu virus example mentioned in Section~\ref{subsec:framework}
also highlights the same fundamental point. Thus, causal mediation
analysis can be uncomfortably far from a completely data-oriented
approach to scientific investigations. It is, however, precisely this
aspect of causal \mbox{mediation} analysis that makes it appealing to those
who resist standard statistical analyses that focus on estimating
treatment effects, an approach which has been somewhat pejoratively
labeled as a ``black-box'' view of causality
(e.g., Skrabanek, \citep{skra94}; Deaton, \citep{deat09}). It may be the case that causal
mediation analysis has the potential to significantly broaden the
scope of statistical analysis of causation and build a bridge between
scientists and statisticians.

There are a number of possible future generalizations of the proposed
methods. First, the sensitivity analysis can potentially be extended
to various nonlinear regression models. Some of this has been done
by Imai, Keele and Tingley \citet{imaikeelting09}. Second, an important generalization
would be to allow multiple mediators in the identification analysis.
This will be particularly valuable since in many applications
researchers aim to test competing hypotheses about alternative causal
mechanisms via mediation analysis. For example, the media framing
study we analyzed in this paper included another measurement (on a
separate group randomly split from the study sample) which was
purported to test an alternative causal pathway. The formal treatment
of this issue will be a major topic of future research. Third,
implications of measurement error in the mediator variable have yet to
be analyzed. This represents another important research topic, as
mismeasured mediators are quite common, particularly in psychological
studies. Fourth, an important limitation of our framework is that it
does not allow the presence of a post-treatment variable that
confounds the relationship between mediator and outcome. As discussed
in Section~\ref{subsec:literature}, some of the previous results avoid
this problem by making additional identification assumptions
(e.g., Robins, \citeyear{robi03}). The exploration of alternative solutions is
also left for future research. Finally, it is important to develop
new experimental designs that help identify causal mediation effects
with weaker assumptions. Imai, Tingley and
Yamamoto \citet{imaitingyama09} present some new
ideas on the experimental identification of causal mechanisms.

\begin{appendix}

\section{\texorpdfstring{Proof of Theorem~\protect\ref{THIDENTIFY}}{Proof of Theorem 1}}
\setcounter{equation}{24}\renewcommand{\theequation}{\arabic{equation}}
\label{app:identify}

First, note that equation~\eqref{eq:YMindepT} in
Assumption~\ref{ASMIGNORABLE} implies
%
\begin{equation}
Y_i(t^\prime,m)  \indep T_i  \vert M_i(t) = m^\prime,\quad X_i = x.
\label{eq:YindepTgivenM}
\end{equation}
%
%
Now, for any $t,t^\prime$, we have
%
\begin{eqnarray} \label{eq:EofY}
&&\quad  \E(Y_i(t, M_i(t^\prime)) \vert X_i = x) \nonumber\\
&&\quad \quad  =  \int\E\bigl(Y_i(t,m) \vert M_i(t^\prime) = m, X_i =
x\bigr)\nonumber\\
&&\quad \hphantom{\quad  =\int\,}
dF_{M_i(t^\prime) \vert X_i = x}(m) \nonumber\\
&&\quad \quad  = \int\E\bigl(Y_i(t,m) \vert M_i(t^\prime) = m, T_i = t^\prime, X_i =
x\bigr) \nonumber\\
&&\quad \hphantom{\quad  =\int\,} dF_{M_i(t^\prime) \mid X_i = x}(m) \nonumber\\
&&\quad \quad  =\int\E(Y_i(t,m) \vert T_i = t^\prime, X_i = x)  \nonumber\\
&&\quad \hphantom{\quad  =\int\,}
dF_{M_i(t^\prime) \mid X_i = x}(m) \nonumber\\
&&\quad \quad  = \int\E(Y_i(t,m) \vert T_i = t, X_i = x)\nonumber
\\
&&\quad \hphantom{\quad  =\int\,} dF_{M_i(t^\prime)
\mid T_i = t^\prime, X_i = x}(m) \nonumber\\
&&\quad \quad  = \int\E\bigl(Y_i(t,m) \vert M_i(t) = m, T_i = t, X_i =
x\bigr)\nonumber\\
&&\quad \hphantom{\quad  =\int\,}
dF_{M_i(t^\prime) \mid T_i = t^\prime, X_i = x}(m) \nonumber\\
&&\quad \quad  = \int\E(Y_i \vert M_i = m, T_i = t, X_i = x)  \nonumber\\
&&\quad \hphantom{\quad  =\int\,} dF_{M_i(t^\prime
) \mid T_i = t^\prime, X_i = x}(m) \nonumber\\
&&\quad \quad  = \int\E(Y_i \vert M_i = m, T_i = t, X_i = x)  \\
&&\quad \hphantom{\quad  =\int\,} dF_{M_i \mid T_i
= t^\prime, X_i = x}(m),\nonumber
\end{eqnarray}
where the second equality follows from
equation~\eqref{eq:YindepTgivenM}, equation~\eqref{eq:YindepMgivenT} is
used to establish the third and fifth equalities,
equation~\eqref{eq:YMindepT} is used to establish the fourth and last
equalities, and the sixth equality follows from the fact that $M_i =
M_i(T_i)$ and $Y_i=Y_i(T_i, M_i(T_i))$. Finally,
equation~\eqref{eq:EofY} implies
\begin{eqnarray*}
&&\E(Y_i(t, M_i(t^\prime)))\\
&&\quad  =   \int
\int\E(Y_i \vert M_i = m, T_i = t, X_i = x) \\
&&\hphantom{\quad=\int\int\,}  dF_{M_i \mid T_i =
t^\prime, X_i = x}(m)  \,  dF_{X_i}(x).
\end{eqnarray*}
 Substituting this expression
into the definition of $\bar\delta(t)$ given by
equations~\eqref{eq:def} and \eqref{eq:defbar} yields the desired
expression for the ACME. In addition, since $\bar\tau= \bar\zeta(t)
+ \bar\delta(t^\prime)$ for any $t,t^\prime=0,1$ and $t\ne t^\prime$
under Assumption~\ref{ASMIGNORABLE}, the result for the average
natural direct effects is also immediate.

\section{\texorpdfstring{Proof of Theorem~\protect\ref{THSEM}}{Proof of Theorem 2}}
\label{app:SEM}\setcounter{equation}{26}\renewcommand{\theequation}{\arabic{equation}}

We first show that under Assumption~\ref{ASMIGNORABLE} the model
parameters in the LSEM are identified.
Rewrite equations~\eqref{eq:MgivenT} and \eqref{eq:YgivenMT} using the
potential outcome notation as follows:
%
\begin{eqnarray} \label
{eq:MgivenT1}
M_i(T_i) & = & \alpha_2 + \beta_2 T_i + \epsilon_{i2}(T_i),\\\label{eq:YgivenMT1}
Y_i(T_i, M_i(T_i)) & = & \alpha_3 + \beta_3 T_i + \gamma M_i(T_i)\nonumber
\\[-8pt]
\\[-8pt] &&{}+
\epsilon_{i3}(T_i,M_i(T_i)),\nonumber
\end{eqnarray}
where the following normalization is used:
$\E(\epsilon_{i2}(t))=\E(\epsilon_{i3}(t,m))=0$ for $t=0,1$ and
$m\in
\mathcal{M}$. Then, equation~\eqref{eq:YMindepT} of
Assumption~\ref{ASMIGNORABLE} implies $\epsilon_{i2}(t)\indep T_i$,
yielding $\E(\epsilon_{i2}(T_i) \vert T_i = t)=\E(\epsilon_{i2}(t))=0$
for any $t=0,1$. Similarly, equation~\eqref{eq:YindepMgivenT} implies
$\epsilon_{i3}(t,m) \indep M_i \vert T_i = t$ for all $t$ and $m$,
yielding $\E(\epsilon_{i3}(T_i, M_i(T_i)) \vert T_i = t, M_i =
m)=\E(\epsilon_{i3}(t,m) \vert T_i=t)=\E(\epsilon_{i3}(t,m))=0$ for any
$t$ and $m$ where the second equality follows from
equation~\eqref{eq:YMindepT}. Thus, the parameters in
equations~\eqref{eq:MgivenT} and \eqref{eq:YgivenMT} are identified
under Assumption~\ref{ASMIGNORABLE}. Finally, under
Assumption~\ref{ASMIGNORABLE} and the LSEM, we can write $\E(M_i
\vert
T_i)=\alpha_2 + \beta_2 T_i$, and $\E(Y_i \vert M_i, T_i)= \alpha_3 +
\beta_3 T_i + \gamma M_i$. Using these expressions and
Theorem~\ref{THIDENTIFY}, the ACME can be shown to equal
$\beta_2\gamma$.

\section{\texorpdfstring{Proof that $\rho=0$ under
Assumption~\protect\ref{ASMIGNORABLE}}{Proof that $\rho=0$ under Assumption 1}}
\label{app:rho}

First, as shown in Appendix~\ref{app:SEM},
Assumption~\ref{ASMIGNORABLE} implies $\E(\epsilon_{i2}(T_i)\vert
T_i)=0$ and $\E(\epsilon_{i3}(T_i, M_i(T_i)) \vert T_i,\break  M_i)=0$ where
the (potential) error terms are defined in
equations~\eqref{eq:MgivenT1} and \eqref{eq:YgivenMT1}. These mean
independence relationships (together with the law of iterated
expectations) imply
\begin{eqnarray*}
0 & = & \E(\epsilon_{i3}(T_i,M_i(T_i))M_i) \\
& = & \E\bigl\{\epsilon_{i3}(T_i,M_i(T_i))\bigl(\alpha_2
+ \beta_2 T_i + \epsilon_{i2}(T_i)\bigr)\bigr\} \\
& = & \E\{\epsilon_{i3}(T_i,M_i(T_i)) \epsilon_{i2}(T_i))\}.
\end{eqnarray*}
Thus, under Assumption~\ref{ASMIGNORABLE}, we have $\rho=0
\Longleftrightarrow\break
\E\{\epsilon_{i2}(T_i)\epsilon_{i3}(T_i,M_i(T_i))\}=0$.

\section{\texorpdfstring{Proof of
Theorem~\protect\ref{THMIDENTIFY1}}{Proof of Theorem 4}}
\label{app:identify1}

First, we write the LSEM in terms of
equations~\eqref{eq:MgivenT} and \eqref{eq:rewrite}. We omit possible
pre-treatment confounders $X_i$ from the model for notational
simplicity, although the result below remains true even if such
confounders are included. Since equation~\eqref{eq:YMindepT} implies
$\E(\epsilon_{ji} \vert T_i) = 0$ for $j=2,3$, we can consistently
estimate $(\alpha_1, \alpha_2, \beta_1,\break  \beta_2)$, where $\alpha_1 =
\alpha_3 + \alpha_2 \gamma$ and $\beta_1 = \beta_3 + \beta
_2\gamma$,
as well as $(\sigma_1^2, \sigma_2^2, \tilde\rho)$. Thus, given a
particular value of $\rho$, we have $\tilde\rho\sigma_1\sigma_2 =
\gamma\sigma_2^2 + \rho\sigma_2\sigma_3$ and $\sigma_1^2 = \gamma^2
\sigma_2^2 + \sigma_3^2 + 2 \gamma\rho\sigma_2\sigma_3$. If $\rho=
0$, then $\gamma= \tilde\rho\sigma_1/\sigma_2$ provided that
$\sigma_3^2 = \sigma_1^2(1-\tilde\rho^2) \ge0$. Now, assume $\rho
\ne0$. Then, substituting $\sigma_3 = (\tilde\rho\sigma_1 -
\gamma\sigma_2)/\rho$ into the\vspace*{1pt} above expression of $\sigma_1^2$ yields
the following quadratic equation: $\gamma^2 -
2\gamma\tilde\rho\sigma_1/\sigma_2 +
\sigma_1^2(\tilde\rho^2-\rho^2)/\{\sigma_2^2(1-\rho^2)\} = 0$.
Solving this equation and using $\sigma_3 \ge0$, we obtain the
following desired expression:\vspace*{-1pt} $\gamma= \frac{\sigma_1}{\sigma_2}
\{
\tilde\rho- \rho\sqrt{(1-\tilde\rho^2)/(1-\rho^2)}\}$. Thus,
given a particular value of $\rho$, $\bar\delta(t)$ is identified.

\section{Nonparametric Sensitivity~Analysis}
\label{app:nonparasens}\setcounter{equation}{28}\renewcommand{\theequation}{\arabic{equation}}

We consider a sensitivity analysis for the simple plug-in
nonparametric estimator introduced in Section~\ref{subsec:nonpar}.
Unfortunately, sensitivity analysis is not as straightforward as the
parametric settings. Here, we examine the special case of binary
mediator and outcome where some progress can be made and leave the
development of sensitivity analysis in a more general nonparametric
case for future research.

We begin by the nonparametric bounds on the ACME without assuming
equation~\eqref{eq:YindepMgivenT} of the sequential ignorability
assumption. In the case of binary mediator and outcome, we can derive
the following sharp bounds using the result of \citet{sjol09}:
%
\begin{eqnarray}\label{eq:sjol1}
&&\max\left\{
\begin{array}{l}
- P_{001} - P_{011} \\
- P_{000} - P_{001} - P_{100} \\
- P_{011} - P_{010} - P_{110}
\end{array}
\right\}\nonumber
\\[-8pt]
\\[-8pt] &&\quad  \leq \bar\delta(1)  \leq
\min\left\{
\begin{array}{l}
P_{101} + P_{111} \\ P_{000} + P_{100} + P_{101} \\
P_{010} + P_{110} + P_{111}
\end{array}
\right\},  \nonumber\\\label{eq:sjol0}
&&\max\left\{
\begin{array}{l}
- P_{100} - P_{110} \\ - P_{001} - P_{100} - P_{101} \\
- P_{110} - P_{011} - P_{111}
\end{array}
\right\}\nonumber
\\[-8pt]
\\[-8pt]
&&\quad  \leq \bar\delta(0)  \leq
\min\left\{
\begin{array}{l}
P_{000} + P_{010} \\ P_{010} + P_{011} + P_{111} \\
P_{000} + P_{001} + P_{101}
\end{array}
\right\},\nonumber
\end{eqnarray}
where $P_{ymt} \equiv\Pr(Y_i = y, M_i = m \vert T_i = t)$ for all
$y,m,\break t \in\{0,1\}$. These bounds always contain zero, implying that
the sign of the ACME is not identified without an additional
assumption even in this special case.

To construct a sensitivity analysis, we follow the strategy of
Imai and Yamamoto \citet{imaiyama10} and first express the second assumption of
sequential ignorability using the potential outcomes notation as
follows:
%
\begin{eqnarray} \label{eq:eqdiff}
&&\quad  \Pr\bigl(Y_i(1,1) = y_{11}, Y_i(1,0) = y_{10},\nonumber\\
&&\quad \phantom{\Pr(} Y_i(0, 1) = y_{01},
Y_i(0,0) = y_{00} \vert M_i = 1, T_i = t^\prime\bigr)\nonumber
\\
&&\quad \quad  =  \Pr\bigl(Y_i(1,1) = y_{11}, Y_i(1,0) = y_{10}, \\
&&\quad \phantom{\quad  =  \Pr(}
Y_i(0, 1) = y_{01},Y_i(0,0) = y_{00}\vert\nonumber\\
&&\quad\hspace*{4pt} \phantom{\quad  =  \Pr(Y_i(1,1) = y_{11}}  M_i = 0, T_i = t^\prime\bigr)\nonumber
\end{eqnarray}
for all $t^\prime, y_{tm}, \in\{0,1\}$. The equality states that
within each treatment group the mediator is assigned independent of
potential outcomes. We now consider the following sensitivity
parameter $\upsilon$, which is the maximum possible difference between
the left- and right-hand side of equation~\eqref{eq:eqdiff}. That is,
$\upsilon$ represents the upper bound on the absolute difference in
the proportion of any principal stratum that may exist between those
who take different values of the mediator given the same treatment
status. Thus, this provides one way to parametrize the maximum
degree to which the sequential ignorability can be violated. (Other,
potentially more intuitive, parametrization are possible, but, as
shown below, this parametrization allows for easier computation of the
bounds.)

Using the population proportion of\vspace*{1pt} each principal stratum, that is,
$\pi_{y_{11}y_{10}y_{01}y_{00}}^{m_1 m_0} \equiv\Pr(Y_i(1, 1) =
y_{11},\break  Y_i(1, 0) = y_{10}, Y_i(0,1) = y_{01}, Y_i(0, 0) = y_{00},
M_i(1) =\break m_1, M_i(0) = m_0)$, we can write this difference as follows:
%
\begin{eqnarray} \label{eq:upsilon1}
&&\hspace*{10pt}\biggl| \frac{\sum_{m_0=0}^1 \pi
^{1m_0}_{y_{11}y_{10}y_{01}y_{00}}}{\sum_{y=0}^1 P_{y11}}-  \frac{\sum_{m_0=0}^1
\pi^{0m_0}_{y_{11}y_{10}y_{01}y_{00}}}{\sum_{y=0}^1 P_{y01}}
\biggr| \nonumber
\\[-8pt]
\\[-8pt]
&&\hspace*{10pt}\quad  \le \upsilon,\nonumber\\[4pt]
\label{eq:upsilon2}
&&\hspace*{10pt}\biggl| \frac{\sum_{m_1=0}^1 \pi
^{m_11}_{y_{11}y_{10}y_{01}y_{00}}}{\sum_{y=0}^1 P_{y10}}-  \frac{\sum_{m_1=0}^1
\pi^{m_10}_{y_{11}y_{10}y_{01}y_{00}}}{\sum_{y=0}^1 P_{y00}} \biggr|\nonumber
\\[-8pt]
\\[-8pt]
&&\hspace*{10pt}\quad
\le \upsilon,\nonumber
\end{eqnarray}
where $\upsilon$ is bounded between $0$ and $1$. Clearly, if and only
if $\upsilon=0$, the sequential ignorability assumption is satisfied.

Finally, note that the ACME can be written as the following linear
function of unknown parame\-ters~$\pi_{y_{11}y_{10}y_{01}y_{00}}^{m_1
m_0}$:
%
\begin{eqnarray} \label{eq:delta.t}
\bar\delta(t)
 &=&  \sum_{m=0}^1 \sum_{y_{1-t,m}=0}^1 \sum_{y_{1,1-m}=0}^1
\sum_{y_{0,1-m}=0}^1\nonumber
\\[-8pt]
\\[-8pt]&& \Biggl(\sum_{m_0=0}^1
\pi_{y_{11}y_{10}y_{01}y_{00}}^{m m_0}
 - \sum_{m_1=0}^1
\pi_{y_{11}y_{10}y_{01}y_{00}}^{m_1 m} \Biggr),\nonumber
\end{eqnarray}
where one of the subscripts of $\pi$ corresponding to $y_{tm}$ is
equal to $1$. Then, given a fixed value of sensitivity parameter
$\upsilon$, you can obtain the sharp bounds on the ACME by numerically
solving the linear optimization problem with the linear constraints
implied by equations~\eqref{eq:upsilon1} and \eqref{eq:upsilon2} as well
as the following relationship implied by the ignorability of the
treatment assignment:
%
\begin{equation}
P_{ymt}  =  \sum_{y_{1-t,m}=0}^1 \sum_{y_{t,1-m}=0}^1
\sum_{y_{1-t,1-m}=0}^1 \sum_{m_{1-t}=0}^1 \pi
_{y_{11}y_{10}y_{01}y_{00}}^{m_1 m_0}
\end{equation}
for each $y,m,t \in\{0,1\}$. In addition, we use the linear
constraint that all $\pi_{y_{11}y_{10}y_{01}y_{00}}^{m_1 m_0}$ sum up
to $1$.

\renewcommand{\thefigure}{\arabic{figure}}
\setcounter{figure}{2}
\begin{figure}[b]

\includegraphics{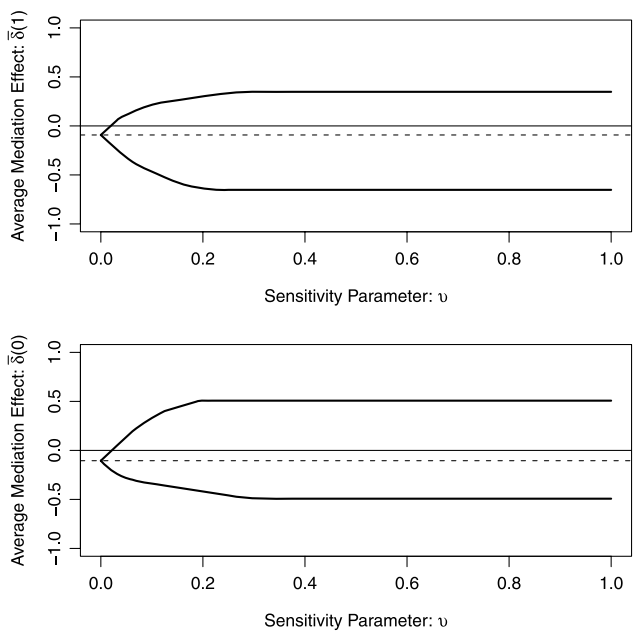}

\caption{Nonparametric sensitivity analysis for the media framing experiment.
In each panel the solid curves show the sharp upper and lower bounds on
the ACME
as a function of the sensitivity parameter $\upsilon$, which
represents the
degree of violation of the sequential ignorability assumption. The horizontal
dashed lines represent the point estimates of $\bar\delta(1)$ (upper
panel) and
$\bar\delta(0)$ (lower panel) under Assumption~\protect\ref{ASMIGNORABLE}.
In contrast
to the parametric sensitivity analysis reported in Section~\protect\ref
{subsec:emp-sens},
the estimates are shown to be rather sensitive to the violation of
Assumption~\protect\ref{ASMIGNORABLE}.} \label{nonpsensfig}
\end{figure}

We apply this framework to the media framing example described in
Sections~\ref{sec:example} and \ref{sec:emp}. For the purpose of illustration,
we dichotomize both the mediator and treatment variables using their
sample medians
as cutpoints. Figure~\ref{nonpsensfig} shows the results of this
analysis. In
each panel the solid curves represent the sharp upper and lower bounds
on the ACME
for different values of the sensitivity parameter $\upsilon$. The horizontal
dashed lines represent the point estimates of $\bar\delta(1)$ (upper
panel) and
$\bar\delta(0)$ (lower panel) under Assumption~\ref{ASMIGNORABLE}.
This corresponds
to the case where the sensitivity parameter is exactly equal to zero
(i.e., $\upsilon=0$),
so that equation~\eqref{eq:eqdiff} holds. The sharp bounds widen as we increase
the value of $\upsilon$, until they flatten out and become equal to
the no-assumption
bounds given in equations~\eqref{eq:sjol1} and \eqref{eq:sjol0}.

The results suggest that the point estimates of the ACME are rather
sensitive to the violation of the sequential ignorability assumption.
For both\vspace*{1pt} $\bar\delta(1)$ and $\bar\delta(0)$, the upper bounds sharply
increase as we increase the value of $\upsilon$ and cross the zero
line at small values of $\upsilon$ [$0.019$ for $\bar\delta(1)$ and
$0.022$ for $\bar\delta(0)$]. This contrasts with the parametric
sensitivity analyses reported in Section~\ref{subsec:emp-sens}, where
the estimates of the ACME appeared quite robust to the violation of
Assumption~\ref{ASMIGNORABLE}. Although the direct comparison is
difficult because of different parametrization and variable coding,
this stark difference illustrates the potential importance of
parametric assumptions in causal mediation analysis; a~significant
part of identification power could in fact be attributed to such
functional form assumptions as opposed to empirical evidence.
\end{appendix}

\section*{Acknowledgments}
A companion paper applying the proposed
methods to various applied settings is available as
Imai, Keele and Tingley \citet{imaikeelting09}. The proposed
methods can be
implemented via the easy-to-use software \texttt{mediation}
(Imai et al. \citep{imaietal10}), which is an R package available at the
Comprehensive R Archive Network
(\href{http://cran.r-project.org/web/packages/mediation}{http://cran.r-project.org/web/packages/mediation}).
The replication data and code for this
article are available for
download as Imai, Keele and
Yamamoto \citet{imaikeelyama10a}. We thank Brian
Egleston, Adam Glynn, Guido Imbens, Gary King, Dave McKinnon,
Judea Pearl, Marc Ratko\-vic, Jas Sekhon, Dustin Tingley, Tyler
VanderWeele and seminar participants at Columbia University,
Harvard University, New York University, Notre Dame, University
of North Carolina, University of Colorado--Boulder, University
of Pennsylvania and University of Wisconsin--Madison for
useful suggestions. The suggestions from the associate editor
and anonymous referees significantly improved the presentation.
Financial support from the National Science Foundation
(SES-0918968) is acknowledged.


\begin{thebibliography}{99}

\bibitem[\protect\citeauthoryear{}{2008}]{albe08}
\textsc{Albert, J.~M.} (2008).
Mediation analysis via potential outcomes models.
\textit{Stat. Med.} \textbf{27}  1282--1304.
\MR{2420158}

\bibitem[\protect\citeauthoryear{}{1996}]{angrimberubi96}
\textsc{Angrist, J.~D.}, \textsc{Imbens, G.~W.} and \textsc{Rubin, D.~B.} (1996).
Identification of causal effects using instrumental variables (with
discussion).
\textit{J. Amer. Statist. Assoc.} \textbf{91} 444--455.

\bibitem[\protect\citeauthoryear{}{2005}]{avinetal05}
\textsc{Avin, C.}, \textsc{Shpitser, I.} and \textsc{Pearl, J.} (2005).
Identifiability of path-specific effects. In
\textit{Proceedings of the International Joint Conference on Artificial
Intelligence}.
Morgan Kaufman, San Francisco, CA.
\MR{2192340}

\bibitem[\protect\citeauthoryear{}{1986}]{barokenn86}
\textsc{Baron, R.~M.} and \textsc{Kenny, D.~A.} (1986).
The moderator--mediator variable distinction in social psychological
research: Conceptual, strategic, and statistical considerations.
\textit{Journal of Personality and Social Psychology} \textbf{51}
1173--1182.

\bibitem[\protect\citeauthoryear{}{1957}]{coch57}
\textsc{Cochran, W.~G.} (1957).
Analysis of covariance: Its nature and uses.
\textit{Biometrics} \textbf{13} 261--281.
\MR{0090952}

\bibitem[\protect\citeauthoryear{}{2009}]{deat09}
\textsc{Deaton, A.} (2009).
Instruments of development: Randomization in the tropics, and the
search for the elusive keys to economic development.
\textit{Proc. Br. Acad.} \textbf{162} 123--160.

\bibitem[\protect\citeauthoryear{}{2006}]{egleetal06}
\textsc{Egleston, B.}, \textsc{Scharfstein, D.~O.}, \textsc{Munoz, B.} and \textsc{West, S.} (2006).
Investigating mediation when counterfactuals are not metaphysical:
Does sunlight {UVB} exposure mediate the effect of eyeglasses on cataracts?
Working Paper 113, Dept. Biostatistics, Johns Hopkins
Univ., Baltimore, MD.

\bibitem[\protect\citeauthoryear{}{2010}]{ellietal10}
\textsc{Elliott, M.~R.}, \textsc{Raghunathan, T.~E.} and \textsc{Li, Y.} (2010).
Bayesian inference for causal mediation effects using principal
stratification with dichotomous mediators and outcomes.
\textit{Biostatistics}.
\textbf{11} 353--372.

\bibitem[\protect\citeauthoryear{}{2009}]{galletal09}
\textsc{Gallop, R.}, \textsc{Small, D.~S.}, \textsc{Lin, J.~Y.}, \textsc{Elliot, M.~R.}, \textsc{Joffe, M.} and \textsc{Ten~Have,
T.~R.} (2009).
Mediation analysis with principal stratification.
\textit{Stat. Med.} \textbf{28}  1108--1130.

\bibitem[\protect\citeauthoryear{}{2007}]{gene07}
\textsc{Geneletti, S.} (2007).
Identifying direct and indirect effects in a non-counterfactual
framework.
\textit{J. Roy. Statist. Soc. Ser.~{B}} \textbf{69}  199--215.
\MR{2325272}

\bibitem[\protect\citeauthoryear{}{2010}]{glyn10}
\textsc{Glynn, A.~N.} (2010).
The product and difference fallacies for indirect effects.
Unpublished manuscript, Dept. Government, Harvard Univ.

\bibitem[\protect\citeauthoryear{}{1960}]{good60}
\textsc{Goodman, L.~A.} (1960).
On the exact variance of products.
\textit{J.~Amer. Statist. Assoc.} \textbf{55}  708--713.
\MR{0117809}

\bibitem[\protect\citeauthoryear{}{2010}]{greehabull10}
\textsc{Green, D.~P.}, \textsc{Ha, S.~E.} and \textsc{Bullock, J.~G.} (2010).
Yes, but what's the mechanism? (don't expect an easy answer).
\textit{Journal of Personality and Social Psychology}
\textbf{98} 550--558.

\bibitem[\protect\citeauthoryear{}{2009}]{hafeschw09}
\textsc{Hafeman, D.~M.} and \textsc{Schwartz, S.} (2009).
Opening the black box: A motivation for the assessment of mediation.
\textit{International Journal of Epidemiology} \textbf{38}  838--845.

\bibitem[\protect\citeauthoryear{}{2010}]{hafevand10}
\textsc{Hafeman, D.~M.} and \textsc{VanderWeele, T.~J.} (2010).
Alternative assumptions for the identification of direct and indirect
effects.
\textit{Epidemiology} \textbf{21}. To appear.

\bibitem[\protect\citeauthoryear{}{2010}]{imaiyama10}
\textsc{Imai, K.} and \textsc{Yamamoto, T.} (2010).
Causal inference with differential measurement error: Nonparametric
identification and sensitivity analysis.
\textit{American Journal of Political Science} \textbf{54}  543--560.

\bibitem[\protect\citeauthoryear{}{2009}]{imaikeelting09}
\textsc{Imai, K.}, \textsc{Keele, L.} and \textsc{Tingley, D.} (2009).
A general approach to causal mediation analysis.
\textit{Psychological Methods}. To appear.

\bibitem[\protect\citeauthoryear{}{2010}]{imaietal10}
\textsc{Imai, K.}, \textsc{Keele, L.}, \textsc{Tingley, D.} and \textsc{Yamamoto, T.} (2010).
Causal mediation analysis using {R}. In
\textit{Advances in Social Science Research Using {R}}
(H. D. Vinod, ed.).
\textit{Lecture Notes in Statist.} \textbf{196} 129--154.
 Springer, New York.

\bibitem[\protect\citeauthoryear{}{2010}]{imaikeelyama10a}
\textsc{Imai, K.}, \textsc{Keele, L.} and \textsc{Yamamoto, T.} (2010).
Replication data for: Identification, inference, and sensitivity
analysis for causal mediation effects.
Available at
\url{http://hdl.handle.net/1902.1/14412}.



\bibitem[\protect\citeauthoryear{}{2009}]{imaitingyama09}
\textsc{Imai, K.}, \textsc{Tingley, D.} and \textsc{Yamamoto, T.} (2009).
Experimental designs for identifying  causal mechanisms.
Technical report, Dept. Politics, Princeton Univ.
Available at \url{http://imai.princeton.edu/research/Design.html}.



\bibitem[\protect\citeauthoryear{}{2003}]{imbe03}
\textsc{Imbens, G.~W.} (2003).
Sensitivity to exogeneity assumptions in program evaluation.
\textit{American Economic Review} \textbf{93}  126--132.

\bibitem[\protect\citeauthoryear{}{2008}]{jo08}
\textsc{Jo, B.} (2008).
Causal inference in randomized experiments with mediational
processes.
\textit{Psychological Methods} \textbf{13}  314--336.



\bibitem[\protect\citeauthoryear{}{2008}]{joffetal08}
\textsc{Joffe, M.~M.}, \textsc{Small, D.}, \textsc{Ten~Have, T.}, \textsc{Brunelli, S.} and \textsc{Feldman, H.~I.} (2008).
Extended instrumental variables estimation for overall effects.
\textit{Int. J. Biostat.} \textbf{4}  Article
4.
\MR{2399287}

\bibitem[\protect\citeauthoryear{}{2007}]{joffetal07}
\textsc{Joffe, M.~M.}, \textsc{Small, D.} and \textsc{Hsu, C.-Y.} (2007).
Defining and estimating intervention effects for groups that will
develop an auxiliary outcome.
\textit{Statist. Sci.} \textbf{22}  74--97.
\MR{2408662}

\bibitem[\protect\citeauthoryear{}{1981}]{juddkenn81}
\textsc{Judd, C.~M.} and \textsc{Kenny, D.~A.} (1981).
Process analysis: Estimating mediation in treatment evaluations.
\textit{Evaluation Review} \textbf{5}  602--619.

\bibitem[\protect\citeauthoryear{}{2008}]{kraekieressekupf08}
\textsc{Kraemer, H.~C.}, \textsc{Kiernan, M.}, \textsc{Essex, M.} and \textsc{Kupfer, D.~J.} (2008).
How and why criteria definig moderators and mediators differ between
the {B}aron \& {K}enny and {M}ac{A}rthur approaches.
\textit{Health Psychology} \textbf{27}  S101--S108.

\bibitem[\protect\citeauthoryear{}{2002}]{kraewilsfairagra02}
\textsc{Kraemer, H.~C.}, \textsc{Wilson, T.}, \textsc{Fairburn, C.~G.} and \textsc{Agras, W.~S.} (2002).
Mediators and moderators of treatment effects in randomized clinical
trials.
\textit{Archives of General Psychiatry} \textbf{59}  877--883.

\bibitem[\protect\citeauthoryear{}{2008}]{mack08}
\textsc{Mac{K}innon, D.~P.} (2008).
\textit{Introduction to Statistical Mediation Analysis}.
Taylor \& Francis, New York.

\bibitem[\protect\citeauthoryear{}{1997}]{nelsclawoxle97}
\textsc{Nelson, T.~E.}, \textsc{Clawson, R.~A.} and \textsc{Oxley, Z.~M.} (1997).
Media framing of a civil liberties conflict and its effect on
tolerance.
\textit{American Political Science Review} \textbf{91}  567--583.

\bibitem[\protect\citeauthoryear{}{2001}]{pear01}
\textsc{Pearl, J.} (2001).
Direct and indirect effects.
In \textit{Proceedings of the Seventeenth
Conference on Uncertainty in Artificial Intelligence}
(J. S. Breese and D. Koller, eds.)
411--420.
Morgan~Kaufman,
San Francisco, CA.

\bibitem[\protect\citeauthoryear{}{2010}]{pear10}
\textsc{Pearl, J.} (2010).
An introduction to causal inference.
\textit{Int. J. Biostat.} \textbf{6}
Article 7.

\bibitem[\protect\citeauthoryear{}{2006}]{petesinilaan06}
\textsc{Petersen, M.~L.}, \textsc{Sinisi, S.~E.} and {van der Laan, M.~J.} (2006).
Estimation of direct causal effects.
\textit{Epidemiology} \textbf{17}  276--284.

\bibitem[\protect\citeauthoryear{}{1999}]{robi99}
\textsc{Robins, J.} (1999).
Marginal structural
models versus structural nested models as tools for causal inference.
In
\textit{Statistical Models in Epidemiology, the Environment and Clinical
Trials} (M.~E. Halloran and D.~A. Berry, eds.)
95--134.
Springer, New York.
\MR{1731682}

\bibitem[\protect\citeauthoryear{}{2003}]{robi03}
\textsc{Robins, J.~M.} (2003).
Semantics of causal {DAG} models and the identification of direct and
indirect effects.
In \textit{Highly Structured Stochastic Systems} (P. J. Green,
N. L. Hjort and S. Richardson, eds.)   70--81. Oxford Univ. Press,
Oxford.
\MR{2082403}

\bibitem[\protect\citeauthoryear{}{1992}]{robigree92}
\textsc{Robins, J.~M.} and \textsc{Greenland, S.} (1992).
Identifiability and exchangeability for direct and indirect effects.
\textit{Epidemiology} \textbf{3}  143--155.

\bibitem[\protect\citeauthoryear{}{2008}]{royetal08}
\textsc{Roy, J.}, \textsc{Hogan, J.~W.} and \textsc{Marcus, B.~H.} (2008).
Principal stratification with predictors of compliance for randomized
trials with 2 active treatments.
\textit{Biostatistics} \textbf{9}  277--289.

\bibitem[\protect\citeauthoryear{}{2004}]{rubi04}
\textsc{Rubin, D.~B.} (2004).
Direct and indirect causal effects via potential outcomes (with
discussion).
\textit{Scand. J. Statist.} \textbf{31}  161--170.
\MR{2066246}

\bibitem[\protect\citeauthoryear{}{2005}]{rubi05}
\textsc{Rubin, D.~B.} (2005).
Causal inference using potential outcomes: Design, modeling,
decisions.
\textit{J. Amer. Statist. Assoc.} \textbf{100} 322--331.
\MR{2166071}

\bibitem[\protect\citeauthoryear{}{2009}]{sjol09}
\textsc{Sj\"{o}lander, A.} (2009).
Bounds on natural direct effects in the presence of confounded
intermediate variables.
\textit{Stat. Med.} \textbf{28}  558--571.

\bibitem[\protect\citeauthoryear{}{1994}]{skra94}
\textsc{Skrabanek, P.} (1994).
The emptiness of the black box.
\textit{Epidemiology} \textbf{5}  5553--5555.

\bibitem[\protect\citeauthoryear{}{1982}]{sobe82}
\textsc{Sobel, M.~E.} (1982).
Asymptotic confidence intervals for indirect effects in structural
equation models.
\textit{Sociological Methodology} \textbf{13} 290--321.

\bibitem[\protect\citeauthoryear{}{2008}]{sobe08}
\textsc{Sobel, M.~E.} (2008).
Identification of causal parameters in randomized studies with
mediating variables.
\textit{Journal of Educational and Behavioral Statistics} \textbf{33}  230--251.

\bibitem[\protect\citeauthoryear{}{2007}]{tenhetal07}
\textsc{Ten~Have, T.~R.}, \textsc{Joffe, M.~M.}, \textsc{Lynch, K.~G.}, \textsc{Brown, G.~K.}, \textsc{Maisto,
S.~A.} and
\textsc{Beck, A.~T.} (2007).
Causal mediation analyses with rank preserving models.
\textit{Biometrics} \textbf{63}  926--934.\
\MR{2395813}

\bibitem[\protect\citeauthoryear{}{2008}]{vand08a}
\textsc{VanderWeele, T.~J.} (2008).
Simple relations between principal stratification and direct and
indirect effects.
\textit{Statist. Probab. Lett.} \textbf{78}  2957--2962.

\bibitem[\protect\citeauthoryear{}{2009}]{vand09}
\textsc{VanderWeele, T.~J.} (2009).
Marginal structural models for the estimation of direct and indirect
effects.
\textit{Epidemiology} \textbf{20}  18--26.

\bibitem[\protect\citeauthoryear{}{2010}]{vand10}
\textsc{VanderWeele, T.~J.} (2010).
Bias formulas for sensitivity analysis for direct and indirect
effects.
\textit{Epidemiology}. To appear.

\bibitem[\protect\citeauthoryear{}{1962}]{zell62}
\textsc{Zellner, A.} (1962).
An efficient method of estimating seemingly unrelated regressions and
tests for aggregation bias.
\textit{J. Amer. Statist. Assoc.} \textbf{57}
348--368.
\MR{0139235}

\end{thebibliography}
\end{document}